\documentclass[a4paper,11pt]{article}
\pdfoutput=1 
\usepackage{jheppub} 
\usepackage[T1]{fontenc} 
\usepackage{comment}
\usepackage{amsmath}
\usepackage{caption}
\usepackage{subcaption}
\usepackage{float}
\usepackage{enumitem}
\usepackage[normalem]{ulem}

\usepackage{cleveref}
\Crefname{figure}{Fig.}{Figs.} 
\Crefname{section}{Sec.}{Secs.}

\newcommand{\Tt}{\tilde{t}}

\newcommand{\gt}{{\tilde{g}}}

\newcommand{\A}{a}

\newcommand{\At}{\tilde{a}}
\newcommand{\Bt}{\tilde{b}}

\newcommand{\py}{\pi}
\newcommand{\pyt}{\tilde{\pi}}

\newcommand{\e}{\varepsilon}
\newcommand{\et}{\tilde{\varepsilon}}
\newcommand{\p}{P}
\newcommand{\pt}{\tilde{P}}
\newcommand{\tp}{\tau_\pi}

\newcommand{\temp}{T_a}
\newcommand{\tempt}{T_b}

\newcommand{\etot}{\mathcal{E}}

\newcommand{\cs}{c_{s}}
\newcommand{\cst}{\tilde{c}_{s}}
\newcommand{\uu}{u}
\newcommand{\uut}{\tilde{u}}

\newcommand{\mn}{{\mu \nu}}

\newcommand{\TM}[1]{{\color{blue}{#1}}}

\title{\boldmath Hybrid thermalization in the large $N$ limit}

\author[a]{Toshali Mitra,}
\author[b]{Sukrut Mondkar,}
\author[c]{Ayan Mukhopadhyay,}
\author[d]{and Alexander Soloviev}

\affiliation[a]{Institute for Theoretical Physics, University of Heidelberg, D-69120 Heidelberg, Germany}
\affiliation[b]{Harish-Chandra Research Institute, A CI of Homi Bhabha National Institute, Chhatnag Road, Jhunsi, Prayagraj (Allahabad) 211019, India}
\affiliation[c]{Instituto de F\'{\i}sica, Pontificia Universidad Cat\'{o}lica de Valpara\'{\i}so,
Avenida Universidad 330, Valpara\'{\i}so, Chile.}
\affiliation[d]{Faculty of Mathematics and Physics, University of Ljubljana, Jadranska ulica 19, SI-1000, Ljubljana, Slovenia}

\emailAdd{t.mitra@thphys.uni-heidelberg.de}
\emailAdd{sukrutmondkar@hri.res.in}
\emailAdd{ayan.mukhopadhyay@pucv.cl}
\emailAdd{alexander.soloviev@fmf.uni-lj.si}

 \abstract{Semi-holography provides a formulation of dynamics in gauge theories involving both weakly self-interacting (perturbative) and strongly self-interacting (non-perturbative) degrees of freedom. These two subsectors interact via their effective metrics and sources, while the full local energy-momentum tensor is conserved in the physical background metric. In the large $N$ limit, the subsectors have their individual entropy currents, and so the full system can reach a pseudo-equilibrium state in which each subsector has a different physical temperature.
 
We first complete the proof that the global thermal equilibrium state, where both subsectors have the \textit{same} physical temperature, can be defined in consistency with the principles of thermodynamics and statistical mechanics.  Particularly, we show that the global equilibrium state is the unique state with maximum entropy in the microcanonical ensemble.
 Furthermore, we show that in the large $N$ limit, a \textit{typical} non-equilibrium state of the full isolated system relaxes to the global equilibrium state when the average energy density is large compared to the scale set by the inter-system coupling. We discuss quantum statistical perspectives.}

\begin{document}
\maketitle

\section{Introduction}

 The mechanism of thermalization of the quark-gluon plasma has been intensely investigated~\cite{Berges:2020fwq, Heinz:2004pj,Strickland:2013uga}, but the most significant questions, such as the nature of the main processes contributing to rapid thermalization and how they can be inferred from experimental data, still require better understanding. To capture the essence of thermalization and other phenomena in such a complex many-body system with quantum gauge fields, it is necessary to combine perturbative approaches with the non-perturbative description of the strongly self-interacting soft sector, especially when the system is effectively in the range of temperatures $T_c- 3 T_c$, where $T_c$ is the deconfinement temperature. 
 
 One such approach is the semi-holographic formalism~\cite{Faulkner:2010tq,Mukhopadhyay:2013dqa,Iancu:2014ava,Doucot:2020fvy,Doucot:2024hzq} which provides such a unified and consistent framework. 
 In this framework, one considers two subsectors described by perturbative and holographic approaches, respectively. These two sectors are coupled via their individual effective metrics in a way that the full energy-momentum tensor is manifestly local and is conserved in the physical background metric (which is Minkowski space for practical purposes). This democratic coupling scheme, proposed in~\cite{Banerjee:2017ozx, Kurkela:2018dku} and reviewed in~\Cref{Sec:SH},  
 can be stated in terms of an effective action and is consistent with the Wilsonian renormalization group. In the large $N$ limit, it has been shown that the dynamics of the full system can be solved explicitly in a self-consistent manner~\cite{Mukhopadhyay:2015smb,Ecker:2018ucc,Mondkar:2021qsf}. 
 
 The phenomenological implications of the semi-holographic approach, {which can be formulated consistently at any level of coarse-graining}, can be derived primarily from the hybrid hydrodynamic attractor surface which was studied explicitly in~\cite{Mitra:2020mei,Mitra:2022uhv} in the context of the Bjorken flow by approximating the dynamical description of each subsector in terms of a M\"uller-Israel-Stewart (MIS) formulation~\cite{Muller:1967zza,Israel:1979wp,Baier:2007ix}. It was shown that not only on the attractor surface, but also for phenomenologically relevant flows (evolving towards the attractor surface) which can be matched to glasma-type approaches~\cite{Kowalski:2007rw} at relevant proper times, the bottom-up thermalization scenario~\cite{Baier:2000sb} is realized (i.e.~it can be demonstrated analytically that the energy density of the hard sector dominates over that of the soft sector) while the full system can be described in terms of a single fluid at late time. Furthermore, the comparison of hydrodynamization of the individual subsectors leads to novel insights into small system collisions. It was shown that although the perturbative sector hydrodynamizes later in Pb-Pb type collisions compared to p-p type collisions, the soft sector can hydrodynamize \textit{earlier} in p-p type collisions compared to Pb-Pb type collisions, both on the hybrid attractor surface and in phenomenologically relevant flows~\cite{Mitra:2022uhv}. This follows from how the hydrodynamization times scale with the total energy density in both subsectors. 

 In this work, we investigate fundamental issues about thermalization in the semi-holographic approach. The two subsystems are isolated in their effective metrics, although they do exchange energy and momentum from the point of view of the physical background metric. In the large $N$ limit, the subsystems also have their individual entropy currents. This implies that, for generic initial conditions, the full system can relax to pseudo-equilibrium states, where each subsystem has a distinct physical temperature, without reaching global equilibrium. In fact, in~\Cref{Sec:DT}, we explicitly demonstrate that the latter can indeed be the case. Furthermore, in the democratic coupling scheme, it is not obvious a priori that the following are satisfied:
 \begin{enumerate}[label=(\roman*)]
     \item we can define a global thermal equilibrium in which the physical temperature of the full system satisfying thermodynamic identities coincides with the equal physical temperatures of the subsystems,
     \item  the entropy density of the full system satisfying thermodynamic identities corresponds to the statistical definition of entropy as the logarithm of the total number of microstates,
     \item  the global thermal equilibrium state is the unique state with maximum entropy in the microcanonical ensemble, and
     \item a typical non-equilibrium isolated state of the full system relaxes to global thermodynamic equilibrium (not a pseudo-equilibrium state) in the limit where the average total energy density is sufficiently large (in consistency with the large $N$ limit).
     \end{enumerate}

 In~\cite{Kurkela:2018dku}, conditions (i) and (ii)  were shown to hold generally in the semi-holographic democratic coupling scheme. We review these proofs in~\Cref{Sec:TSC} while stating the issue of pseudo-equilibrium states explicitly and extending the concept of statistical entropy to the latter. In \Cref{Sec:GEM} we prove that the condition (iii) mentioned above is indeed satisfied, i.e., the global equilibrium state, which can be consistently defined, is indeed the unique maximum entropy state of the micro-canonical ensemble. 

 {The formulation of semi-holography allows us to study the dynamics of the full system at any level of coarse-graining. We will address the fundamental issues in thermalization by employing effective MIS-type descriptions of the two subsystems in which a microstate configuration of the full system is specified phenomenologically by the sub-system energy-momentum tensor themselves. This allows us to study thermalization by following the evolution of the coarse-grained (phenomenological) entropies of the sub-systems and the full system as elaborated in Sec. \ref{Sec:Micro} via the example of two coupled effective kinetic theories.}

 In Sec.~\ref{Sec:DT}, we employ the coarse-grained MIS description for both sectors and demonstrate that when the average total energy density is large, a \textit{typical} non-equilibrium state of the full isolated system does indeed relax to the global equilibrium state, by formulating this criterion in precise mathematical terms. Finally, in Sec.~\ref{Sec:Conc}, we compare our results with quantum statistical approaches such as the eigenstate thermalization hypothesis, and discuss further studies which are necessary for phenomenological applications.

\begin{figure}[h]
    \centering
    \includegraphics[width=0.5\linewidth]{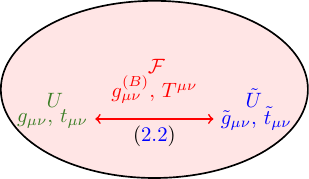}
    \caption{The isolated system, $\mathcal{F}$, is composed of two subsystems, $\mathcal{U}$ and $\tilde{\mathcal{U}}$, which individually satisfy their own conservation equations \cref{Eq:WI} in their respective background metric. The two subsystems are coupled and interact via deformations of their metric, given by \cref{Eq:metriccoup}. }
    \label{fig:micro-picture}
\end{figure}

\section{Semi-holography}\label{Sec:SH}

Semi-holography is an effective framework that incorporates both the perturbative and non-perturbative degrees of freedom consistently for a wide range of energy scales. The term "semi-holography" was coined by Faulkner and Polchinski \cite{Faulkner:2010tq} in the context of holographic non-Fermi liquid models, in which they coupled a dynamical boundary field to IR-CFT.~\footnote{See~\cite{Mukhopadhyay:2013dqa,Iancu:2014ava,Doucot:2020fvy,Doucot:2024hzq} for more on the semi-holographic approach to non-Fermi liquids.} Later, this formulation was implemented in heavy ion collisions in~\cite{Iancu:2014ava} where a classical Yang-Mills field describing the weakly coupled gluon modes is coupled to the strongly coupled conformal field theories, which has a gravity dual. 

The semi-holographic framework proposes a democratic formulation~\cite{Banerjee:2017ozx} in which at any energy scale, the full theory is defined by both the weakly coupled perturbative sectors as well as the strongly coupled non-perturbative sector, which has a classical gravity dual in the large $N$ limit. The two sectors, otherwise isolated from each other, are allowed to interact by deforming the marginal and relevant couplings of each sector, which get promoted to local algebraic functions of the corresponding operators, and this can be implemented via an effective action~\cite{Banerjee:2017ozx,Kurkela:2018dku}. This coupling scheme ensures the renormalizability of the effective action of the full system in the large $N$ limit. The democratic coupling also allows for the construction of low-energy dynamics of the full system from the effective low-energy descriptions of the individual sectors. The two sectors living in their respective effective metric backgrounds are assumed to share the same topological space so that we can describe the full system, including these sectors, using the same coordinates. Moreover, the energy-momentum tensor of the full system is conserved in the physical background metric while the subsectors are closed in their respective background metrics and are allowed to exchange energy and momentum with respect to the physical background metric. This picture is summarized in~\Cref{fig:micro-picture}.

The dynamics of the full system is solved by solving the dynamics of the individual sectors iteratively in a self-consistent manner \cite{Iancu:2014ava}. The iterative method specifies the initial conditions for the dynamical variables of each sector, including the boundary condition of the bulk field. At each iteration, the initial conditions are kept fixed, and the individual sectors are solved independently with their effective metrics and couplings determined self-consistently. When the iteration converges, the energy-momentum tensor of the full system becomes conserved in the physical background metric. Convergence of the dynamics of the full system has been confirmed in various numerical investigations where it has been shown that the convergence is highly non-trivial~\cite{Mukhopadhyay:2015smb,Ecker:2018ucc}. This iterative method with proper initial conditions for heavy-ion collisions has been described in~\cite{Iancu:2014ava}.

As previously stated, the democratic formulation allows for the coarse-grained description of the entire system from the effective coarse-grained dynamics of individual sectors.
This allows for phenomenological construction, where the stress tensor of the entire system is a polynomial of the stress tensors of the subsectors. Thus, a hydrodynamic description of the individual subsectors is sufficient to develop the same for the full physical system. However, this property is unique to democratic coupling in a semi-holographic framework. Otherwise, in systems such as water-oil mixtures, we must know the nature of intermolecular interactions in addition to the hydrodynamic descriptions of water and oil to define the mixture using hydrodynamic theory. To better understand the preceding discussion, we consider the example of metric coupling (rank-2 tensor)~\cite{Kurkela:2018dku}, which is relevant for the following section. 

\subsection{Effective metric coupling}

Consider a dynamical system $ \mathcal{F}$ in a fixed metric background $ g^{(B)}_\mn $. The physical system is composed of two sub-sectors $ \mathcal{U} $ and $ \tilde{\mathcal{U}} $, which correspond to the perturbative and non-perturbative degrees of freedom, see~\Cref{fig:micro-picture}. The two sub-sectors live in their respective metric background $ g_\mn$ and $ \tilde{g}_\mn $. The effective action of the system that captures the deformation of the marginal/relevant coupling of the individual sectors and their interaction is~\cite{Kurkela:2018dku}
\begin{eqnarray} \label{Eq:semi_action}
S_{tot}[\phi, \tilde{\phi},g_\mn,\tilde{g}_\mn, g^{(B)}_\mn] &=& {S}[\phi,g_\mn] 
+ \tilde{{S}}[\tilde{\phi},\tilde{g}_\mn] \nonumber\\ &+& \int d^{d} x ~ \Big[ \frac{1}{2 \gamma} ~ \sqrt{-g^{(B)}}~ tr[(g-g^{(B)})\cdot {g^{(B)}}^{-1} \cdot(\tilde{g} - g^{(B)})\cdot {g^{(B)}}^{-1}] \nonumber \\
&+& \frac{\gamma'}{2 \gamma }  ~ ~\sqrt{-g^{(B)}} ~ \frac{(tr[g\cdot {g^{(B)}}^{-1}] - d ) (tr[\tilde{g}\cdot {g^{(B)}}^{-1}]-d)}{ \gamma'd-\gamma}  \Big]
\end{eqnarray}
in $d$-dimensional spacetime. ${S} $ and $\tilde{{S}} $ are the effective actions of the respective sectors, $ \mathcal{U}$ and $\tilde{\mathcal{U}}$ and $\phi $ and $ \tilde{\phi}$ denote all the matter fields in the two sub-sectors. $ \gamma$ and $ \gamma '$ are the dimensionful semi-holographic coupling of mass dimension $ - d$ whose ratio gives the dimensionless coupling $ r := -\frac{\gamma'}{\gamma}$.
The effective metrics $ g_\mn$ and $\tilde{g}_\mn$ act as auxiliary fields and capture the marginal deformation of the respective sectors. As mentioned earlier, the deformation is taken care of by promoting the effective metrics as local algebraic functions of the operators of the sub-sectors. The algebraic equations of the effective metric are obtained by varying the action (\ref{Eq:semi_action}) with respect to the individual effective metrics. The lowest-order metric coupling equations that dictate the coupling between the sub-sectors explicitly are
\begin{align} \label{Eq:metriccoup}
g_\mn &= g^{(B)}_\mn + \frac{\sqrt{- \tilde{g}}}{\sqrt{- g^{(B)}}}
\Big[ \gamma ~ \tilde{t}^{\alpha \beta} g^{(B)}_{\alpha \mu} g^{(B)}_{\beta \nu}   + \gamma' ~ tr(\tilde{t}\cdot g^{(B)}) g^{(B)}_\mn  \Big],\nonumber \\
\tilde{g}_\mn &= g^{(B)}_\mn + \frac{\sqrt{- {g}}}{\sqrt{- g^{(B)}}}
\Big[ \gamma ~ {t}^{\alpha \beta} g^{(B)}_{\alpha \mu} g^{(B)}_{\beta \nu}   + \gamma' ~ tr({t}\cdot g^{(B)}) g^{(B)}_\mn  \Big],
\end{align}
where $ t^{\alpha \beta}$ and $\tilde{t}^{\alpha \beta}$ are the energy-momentum tensor of the respective subsectors
\begin{eqnarray}
t^{\alpha \beta} = -\frac{2}{\sqrt{-g}} \frac{\delta {S}}{\delta g_{\alpha \beta}}, \hspace{1cm} \tilde{t}^{\alpha \beta} = -\frac{2}{\sqrt{-\tilde{g}}} \frac{\delta \tilde{{S}}}{\delta\tilde{g}_{\alpha \beta}}.
\end{eqnarray}
Given that the subsystem actions $S$ and $\tilde{S}$ are diffeomorphism invariant, the sub-sector energy-momentum tensors are conserved with respect to their effective metric backgrounds, i.e., they satisfy the Ward identities, 
\begin{align} \label{Eq:WI}
\nabla_\alpha t^{\alpha \beta} = 0, \hspace{1cm} \tilde{\nabla}_\alpha \tilde{t}^{\alpha \beta} = 0,
\end{align}
where $ \nabla$ and $ \tilde{\nabla}$ are the covariant derivatives with respect to the effective metrics.  Furthermore, the energy-momentum tensor of the full system $\mathcal{F}$ can be obtained as a polynomial of the energy-momentum of the individual sectors by varying the action (\ref{Eq:semi_action}) with respect to the background metric $ g^{(B)}_\mn$
\begin{eqnarray} \label{Eq:fullem}
T^{\mn} &:=& -\frac{2}{\sqrt{-g^{(B)}}} \frac{\delta {S}_{tot}}{\delta g^{(B)}_\mn},\nonumber \\
T^\mu_\nu &=& \frac{1}{2} \Bigg( (t^\mu_\nu + t^\mu_\nu) \frac{\sqrt{-g}}{\sqrt{-g^{(B)}}} + (\tilde{t}^\mu_\nu + \tilde{t}^\mu_\nu) \frac{\sqrt{-\tilde{g}}}{\sqrt{-g^{(B)}}} \Bigg) + \Delta K \delta^\mu_\nu
\end{eqnarray}
where
\begin{eqnarray}\label{Eq:fullem1}
\Delta K = &-& \frac{\gamma}{2} ~ \Bigg( t^{\rho \alpha} \frac{\sqrt{-g}}{\sqrt{-g^{(B)}}} \Bigg) ~ g_{\alpha \beta}^{(B)} ~ \Bigg( \tilde{t}^{\beta \sigma} \frac{\sqrt{-\tilde{g}}}{\sqrt{-g^{(B)}}} \Bigg) ~ g^{(B)}_{\sigma \rho} \nonumber \\
&-& \frac{\gamma'}{2} ~ \Bigg( t^{\alpha \beta} \frac{\sqrt{-g}}{\sqrt{-g^{(B)}}} \Bigg) ~ g_{\alpha \beta}^{(B)} ~ \Bigg( \tilde{t}^{\sigma \rho} \frac{\sqrt{-\tilde{g}}}{\sqrt{-g^{(B)}}} \Bigg) ~ g^{(B)}_{\sigma \rho}.
\end{eqnarray}
We note that the full energy-momentum tensor is explicitly a local algebraic functional of the subsystem energy-momentum tensors. We can readily check that the full energy-momentum tensor $ T^\mu_\nu$ is conserved in the physical background $ g^{(B)}_\mn$ provided that the coupling equations~\eqref{Eq:metriccoup} and the sub-sector Ward identities~\eqref{Eq:WI} are satisfied, i.e.,~\eqref{Eq:metriccoup} and~\eqref{Eq:WI} imply that
\begin{eqnarray} \label{Eq:TEM}
\nabla_{\mu}^{(B)} T^\mu_\nu = 0.
\end{eqnarray}
As the full action is not invariant under individual diffeomorphisms of the subsystems but only under their diagonal combination, the Ward identities of the individual subsystems imply the Ward identity for the full system.

The above formalism simplifies in the large $N$ limit where expectation values of multi-trace gauge-invariant operators factorize. In this case, the coupling equations~\eqref{Eq:metriccoup} give well-defined non-fluctuating effective metrics of the subsystems in terms of the expectation values of the energy-momentum tensors of the subsectors in any state.

\subsection{On microstates and coarse-grained entropy in semi-holography}\label{Sec:Micro}

One of the main advantanges of the democratic effective couplings described above is that microstates of the full system can be defined in the semi-holographic framework at any level of coarse-graining in the large $N$ limit. We will not pursue this explicitly here in the full quantum field theory although this follows from the action formulation \eqref{Eq:semi_action}. Assuming that both the hard (UV) and soft (IR) sectors are weakly coupled, we can use effective kinetic theories to describe each sector. Here we illustrate that we can define the microstates and their time-evolution in this case as the energy-momentum tensor and effective metric of each sub-sector is well-defined self-consistently together in terms of the microscopic variables, namely the particle distribution functions. This can be generalized to other contexts, e.g. if we replace the kinetic description of the soft (IR) sector with a holographic description in terms of a classical theory of gravity in one higher dimension.\footnote{In this case, the initial condition of the holographic sector is stated in terms of gravitational initial conditions. For an illustration see \cite{Ecker:2018ucc,Mondkar:2021qsf} in which classical fields describing the UV sector is coupled with a holographic IR sector.}

Consider the two sub-systems $\mathcal{U}$ and $\tilde{\mathcal{U}}$ to be gases of weakly interacting massless particles living in their respective background metrics, $g_{\mu\nu}$ and $\tilde{g}_{\mu\nu}$. Let us assume that we can employ effective kinetic theory descriptions for each of these sub-systems. For each particle in sub-system $\mathcal{U}$, the on-shell four momentum $p^\mu$ should satisfy $ g_{\mu\nu} p^\mu p^\nu = 0$ and similarly in sub-system $\tilde{\mathcal{U}}$, the on-shell four-momentum of each massless particle satisfies $\tilde{g}_{\mu\nu} \tilde{p}^\mu \tilde{p}^\nu = 0$. Clearly, we can solve the on-shell conditions to obtain the on-shell effective energies $p^0(x)$ and $\tilde{p}^0(x)$ as functions of $p^i$ and $g_{\mu\nu}(x)$, and $\tilde{p}^i$ and $\tilde{g}_{\mu\nu}(x)$, respectively. The microstates of $\mathcal{U}$ and $\tilde{\mathcal{U}}$ are specified by the particle distribution functions $f(x, p^i)$ and $\tilde{f}(x, \tilde{p}^i)$, respectively at a given point of time $x^0$.

The distribution functions allow us to compute the macroscopic quantities, particularly the energy-momentum tensor $t^{\mu\nu}$ and $\tilde{t}^{\mu\nu}$ for each sector,
\begin{align}
   & t^{\mu\nu}(x) = \int \frac{d^3 p}{ (2\pi)^3 \sqrt{-g(x)}} \frac{p^\mu(x) p^\nu(x)}{p^0(x)} f(x, p^i),\nonumber \\
    &\tilde{t}^{\mu\nu}(x) = \int \frac{d^3 \tilde{p}}{ (2 \pi)^3 \sqrt{-\tilde{g}(x)}} \frac{\tilde{p}^\mu(x) \tilde{p}^\nu(x)}{\tilde{p}^0(x)} \tilde{f}(x, \tilde{p}^i).
\end{align}
Note that $t^{\mu\nu}(x)$ and $\tilde{t}^{\mu\nu}(x)$ should be obtained self-consistently along with the effective metrics $g_{\mu\nu}(x)$ and $\tilde{g}_{\mu\nu}(x)$ by solving the algebraic coupling equations \eqref{Eq:metriccoup}. All of these can be determined locally given the microscopic variables, $f(x, p^i)$ and $\tilde{f}(x, \tilde{p}^i)$.
 %we obtain two sets of Boltzmann equations, one for each of the sub-sectors, characterised by their respective distribution functions $f$ and $\tilde{f}$, and relaxation time $\tau_R$ and $\tilde{\tau}_R$. 
%The flow velocity for each sub-sectors is defined to be timelike, i.e., $u_\mu u^\mu = -1$ and $\tilde{u}_\mu \tilde{u}^\mu = -1$, and the four-momentum of each sector satisfies the on-shell condition expressed as $ g_{\mu\nu} p^\mu p^\nu = 0$, and  $\tilde{g}_{\mu\nu} \tilde{p}^\mu \tilde{p}^\nu = 0$. The corresponding equilibrium distribution function for the sub-sectors with effective temperature $T_1$ and $T_2$ reads,

It could be useful to see how one can define the kinetic equations for each sub-sector in the RTA (relaxation time approximation). First, we need to define local equilibrium particle distributions via matching conditions (see e.g.~\cite{Romatschke:2015gic,Bajec:2024jez,Bajec:2025dqm}). To do so, we construct the flow vectors $u^\mu$ and $\tilde{u}^\mu$ in the Landau frame so that these are time-like eigenvectors of $t^\mu_{\,\nu}$ and $\tilde{t}^\mu_{\,\nu}$, respectively, and they satisfy $u^\mu g_{\mu\nu}u^\nu = \tilde{u}^\mu \tilde{g}_{\mu\nu}\tilde{u}^\nu = -1$. The energy densities of $\mathcal{U}$ and $\tilde{\mathcal{U}}$ are
\begin{align}
    &\varepsilon = \int \frac{d^3 p}{ (2\pi)^3 \sqrt{-g}} \frac{(u_\mu(x) p^\mu(x))^2}{p^0(x)} f(x, p^i), \nonumber\\
    &\tilde{\varepsilon} = \int \frac{d^3 \tilde{p}}{ (2\pi)^3 \sqrt{-g}(x)} \frac{(u_\mu(x) p^\mu(x))^2}{p^0(x)} f(x, p^i), 
\end{align} 
respectively. Finally, we can define the local equilibrium particle distributions 
\begin{align}
  f_{eq} = \frac{1}{e^{-p^\mu u^\nu g_{\mu\nu}/T_1}-1}, \,\,\,\,\,\,\, \tilde{f}_{eq} =  \frac{1}{e^{-\tilde{p}^\mu \tilde{u}^\nu \tilde{g}_{\mu\nu}/T_2}-1}. \label{Eq:kineticdistributionfunc2}
\end{align}
where $T_1$ and $T_2$ are defined from $\varepsilon(T_1)$ and $\tilde\varepsilon(T_2)$, respectively, via the equations of states of the respective subsystems. %Finally, in terms of the covariant particle distribution of $\mathcal{U}$ given by
% \begin{equation}
%     F(x, p^\mu) = f(x, p^i)\delta(p^\mu g_{\mu\nu}p^\nu)\theta(p^0)
% \end{equation}
%we can define the RTA Boltzmann equation
% \begin{align}
%    p^\mu \partial_\mu F + p_\mu p^\nu \Gamma^\mu_{\alpha \nu} \frac{\partial F}{\partial p_\alpha} = \frac{p^\mu u_\mu }{\tau_R} (F-F_{eq}) \label{Eq:kineticeq}
% \end{align}
The evolution of the one-particles distribution function is given by the RTA Boltzmann equation
\begin{align}
   p^\mu \partial_\mu f + p^\mu p^\nu \Gamma^i_{\mu \nu} \frac{\partial f}{\partial p_i} = \frac{p^\mu u_\mu }{\tau_R} (f-f_{eq}) \label{Eq:kineticeq}
\end{align}
for $\mathcal{U}$ with $\tau_R = C_\tau T_1^{-1}$ and $C_\tau$ a constant. Similarly, we can construct the RTA Boltzmann equation for $\tilde{\mathcal{U}}$ with $\tilde\tau_R = \tilde{C}_\tau T_2^{-1}$ and $\tilde{C}_\tau$ a constant. 

It is not hard to see that the full dynamics can be solved self-consistently with initial conditions where $f(x,p^i)$ and $\tilde{f}(x, \tilde{p}^i)$ are given at any point of time. We will explicitly describe an algorithm in a simplified context in this paper where each subsector is further simplified with a BRSSS description as in \cite{Mitra:2022uhv}. The interested reader can study another illustration of semi-holographic dynamics in which a quantum harmonic oscillator is coupled to a quantum dot which admits a holographic dual description in terms of a two-dimensional gravitational theory \cite{Kibe:2023ixa}.

Our illustration with coupled effective kinetic theories also demonstrates that just like the full dynamics we can also define coarse-grained entropies of the subsystems at any level of coarse-graining. In case of the full quantum theory, the coarse-grained entropy of a state is defined as the von-Neuman entropy of the density matrix which has the maximum von-Neuman entropy in the set of all density matrices which reproduce the mean values of simple observables in the given state. In the context of the effective kinetic descriptions, we can define the entropy currents of the (bosonic) sub-systems via
\begin{align} \label{Eq:entropy}
   & s^\mu = g_d\int \frac{d^3 p}{(2 \pi)^3  \sqrt{-g}} \frac{p^\mu}{p^0} [f \ln f - (1+f)\ln(1+f) ],\nonumber \\
   & \tilde{s}^\mu = \tilde{g}_d \int \frac{d^3 \tilde{p}}{(2 \pi)^3 \sqrt{-\tilde{g}}} \frac{\tilde{p}^\mu }{\tilde{p}^0}\tilde{[f} \ln\tilde{f} - (1+\tilde{f})\ln(1+\tilde{f}) ] .
   %& \nabla_\mu s^\mu = \int \frac{d^3 p}{ (2 \pi)^3 p^0 \sqrt{-g}} \frac{p^\mu u_\mu }{\tau_R} (f-f_{eq}) \ln f
\end{align}
where $g_d$ and $\tilde{g}_d$ are the respective degeneracy factors. If the kinetic description of the soft sub-system is replaced by a holographic description, a phenomenological entropy current can be derived from the apparent horizon (see e.g.~\cite{Booth:2010kr,Jansen:2020ign}). 

To understand fundamental issues about thermalization, we will employ effective BRSSS descriptions of the perturbative and non-perturbative subsystems assuming them to be conformal (note that the effective metric coupling breaks conformality). In this context the microstates of the sub-systems are specified by their energy-momentum tensors ($t^{\mu}_{\,\nu}$ and $\tilde{t}^{\mu}_{\,\nu}$) themselves instead of fine-grained variables such as the particle distribution functions. In this case, the microstates and the entropies are defined phenomenologically. This simplification will allow us to study the evolution of the coarse-grained (phenomenological) entropies of the sub-systems and the full system, and thus thermalization.

%\AS{Quick decoupling discussion here}

%\AS{Entropy (and its production) $\int f \log f+ \int \tilde{f} \log \tilde{f}$. Discuss entropy current growth in 2 gas example following argument, see eq. 3.11 1805.05213\\
%Small computation for $f=f(t,p)$}

\section{Thermodynamic and statistical consistencies, and the problem of pseudo-equilibria}\label{Sec:TSC}

\subsection{The global equilibrium}

A crucial test of the democratic effective metric coupling is its consistency with thermodynamics and statistical mechanics. As shall be described below, in thermal equilibrium, basic principles of thermodynamics and statistical mechanics determine the local temperature and entropy densities of the full system in terms of those of the subsystems. The democratic coupling scheme determines the full energy-momentum tensor, and therefore the total energy density and pressure given the physical temperature and the equations of state of the individual subsystems. The total energy density and pressure can then be used to obtain the local temperature and entropy densities of the full system via thermodynamic identities. If the democratic coupling system is consistent physically, then the local temperature and entropy densities of the full system determined in these two different ways must agree. In this section, we first elaborate on these requirements of thermodynamic and statistical consistencies, and then review the proof that these are indeed satisfied by the democratic effective metric coupling scheme.

For thermal equilibrium to exist in any background metric, a global time coordinate $\tau$ must exist. The vector field $\xi^\mu\partial_\mu =\partial_\tau$ should be a hypersurface orthogonal Killing vector field in the physical static background metric $g^{(B)}_{\mu\nu}$, and also in the two effective static background metrics $g_{\mu\nu}$ and $\tilde{g}_{\mu\nu}$, in which the two subsystems live. Assume that the two subsystems can reach thermal equilibrium at constant physical temperature $\mathcal{T}_0$, which is independent of space and time. The two subsystems generically \textit{should} have two \textit{different effective local} temperatures $T_1$ and $T_2$ so that the proper lengths of their (local) Euclidean time circles are $T_1^{-1}$ and $T_2^{-1}$, respectively, with equal periods set to $\beta =\mathcal{T}_0^{-1}$ globally. The local temperature $\mathcal{T}$ of the full system should be the inverse of the proper length of the Euclidean time circle measured in the physical background metric $g^{(B)}_{\mu\nu}$ with the period globally set to $\beta=\mathcal{T}_0^{-1}$. Therefore,
\begin{align}\label{Eq:TC1}
    \mathcal{T}^{-1}(x) &= \int_0^{\beta} \sqrt{- g^{(B)}_{\tau\tau}(x)}\,{\rm d}\tau,\\  T_1^{-1}(x) &= \int_0^{\beta} \sqrt{- g_{\tau\tau}(x)}\, {\rm d}\tau, \\ T_2^{-1}(x) &= \int_0^{\beta} \sqrt{- \tilde{g}_{\tau\tau}(x)} \,{\rm d}\tau,
\end{align}
where we have made the dependence on the spatial coordinates $x$ manifest. Since $\partial_\tau$ is a Killing vector field,  $g^{(B)}_{\tau\tau}$, $g_{\tau\tau}$ and $\tilde{g}_{\tau\tau}$ are independent of $\tau$. Therefore, 
\begin{equation}\label{Eq:TC}
   \sqrt{- g^{(B)}_{\tau\tau}(x)} \,\mathcal{T}(x) =\sqrt{-g_{\tau\tau}(x)}\, T_1(x)= \sqrt{-\tilde{g}_{\tau\tau}(x)}\, T_2(x) = \beta^{-1}=\mathcal{T}_0 .
\end{equation}

Furthermore, in the democratic effective metric coupling scheme, the two subsystems interact only by mutually deforming their effective metric backgrounds. They are thus isolated from each other in their self-consistent effective metric backgrounds (although they exchange energy and momentum from the point of view of the physical background metric). Therefore, the total number of microstates of the full system in thermal equilibrium should just be the products of the number of microstates of the subsystems, and consequently the entropy density $\mathcal{S}$ of the full system should just be the sum of the subsystem entropy densities normalized by their effective spatial volumes. Explicitly, we should have
\begin{equation}\label{Eq:SC}
    \mathcal{S}(x)\sqrt{g^{(B)}_\perp(x)} = s_1(x) \sqrt{g_\perp(x)} + s_2(x) \sqrt{\tilde{g}_\perp(x)},
\end{equation}
where $\sqrt{g^{(B)}_\perp}$, $\sqrt{g_\perp}$ and $\sqrt{\tilde{g}_\perp}$ are the volume elements of the spatial hypersurfaces orthogonal to the global time-like Killing vector, $\xi^\mu$, in the respective static metric backgrounds; $\mathcal{S}$ is the total entropy density, and $s_1$ and $s_2$ are the entropy densities of the corresponding subsystems. 

We can readily see that given the (constant) physical temperature $\mathcal{T}_0$ and the equations of states of the two subsystems ($P(\e)$ and $\tilde P(\tilde\e)$), we can determine the energy-momentum tensor of the full system (for an illustration, see next subsection), and thus the total energy density $\mathcal{E}$ and the total pressure $P$. 
The local temperature $\mathcal{T}$ of the full system and its entropy density $\mathcal{S}$ can then be obtained from the thermodynamic identities
\begin{eqnarray}\label{Eq:thermoidentities-full}
\mathcal{E} + \mathcal{P} = \mathcal{S} \mathcal{T}, \hspace{1cm}  d \mathcal{E} = \mathcal{T} d \mathcal{S}.
\end{eqnarray}
The thermodynamic identities for the subsystems are
\begin{eqnarray}\label{Eq:thermoidentities}
\e + P = T_1 s_1, \hspace{1cm} {\rm d}\e  = T_1 {\rm d}s_1,\nonumber\\
\et + \tilde{P} = T_2 s_2, \hspace{1cm} {\rm d}\et = T_2 {\rm d}s_2,
\end{eqnarray}
where $ \e(\et)$ and $ P (\tilde{P})$ are the energy density and pressure of the respective subsystems.
The question is whether $\mathcal{T}$ and $\mathcal{S}$ determined via~\eqref{Eq:thermoidentities-full} agree with~\eqref{Eq:TC} and~\eqref{Eq:SC} as required by consistency with basic principles of thermodynamics and statistical mechanics.  

\subsection{Proof of thermodynamic/statistical consistencies}
The proof of thermodynamic and statistical consistencies can be obtained by considering the full system $ \mathcal{F} $ living in a $d-$dimensional isotropic static spacetime metric characterized by the static gravitational potential $\phi(x)$ (not a function of time) so that
\begin{eqnarray} \label{Eq:backgroundmetric}
g_\mn^{(B)} = \text{diag}(-e^{-2\phi(\bold{x})},\underbrace{1, \cdots,1}_{d-1}).
\end{eqnarray}
The effective metrics of the two subsystems $\mathcal{U}$ and $\tilde{\mathcal{U}}$ should also be static and isotropic, and therefore assume the forms 
\begin{eqnarray}\label{Eq:Effmet}
g_\mn = \text{diag}\Big(-a(\bold{x})^2, \underbrace{b(\bold{x})^2, \cdots, b(\bold{x})^2}_{d-1}\Big), \quad \tilde{g}_\mn = \text{diag}\Big(-\tilde{a}(\bold{x})^2, \underbrace{\tilde{b}(\bold{x})^2, \cdots, \tilde{b}(\bold{x})^2}_{d-1}\Big).
\end{eqnarray}
For the full system to be in thermal equilibrium, the subsystems should be in local equilibrium at local temperatures $T_1$ and $T_2$ so that the subsystem energy-momentum tensors should assume the forms
\begin{eqnarray} \label{Eq:subsystemEM}
t^\mn = \text{diag}\Bigg( \frac{\e(T_1(\bold{x}))}{a(\bold{x})^2}, \underbrace{\frac{P(T_1(\bold{x}))}{b(\bold{x})^2}, \cdots, \frac{P(T_1(\bold{x}))}{b(\bold{x})^2}}_{d-1} \Bigg), \\
\tilde{t}^\mn = \text{diag}\Bigg( \frac{\et(T_2(\bold{x}))}{\tilde{a}(\bold{x})^2}, \underbrace{\frac{\tilde{P}(T_2(\bold{x}))}{\tilde{b}(\bold{x})^2}, \cdots, \frac{\tilde{P}(T_2(\bold{x}))}{\tilde{b}(\bold{x})^2}}_{d-1} \Bigg).
\end{eqnarray} For the subsystem energy-momentum tensors to be conserved locally in their effective background metrics, we have
\begin{eqnarray}\label{Eq:subsystemWI}
\frac{\partial_i P}{\e+P} + \frac{\partial_i a}{a} = 0, \hspace{1cm} \frac{\partial_i \tilde{P}}{\et+\tilde{P}} + \frac{\partial_i \tilde{a}}{\tilde{a}} = 0
\end{eqnarray}
where $ \partial_i $ denotes the derivative w.r.t.~a spatial coordinate. Using the thermodynamic identities~\eqref{Eq:thermoidentities} which imply that ${\rm d}P = s_1 {\rm d}T_1$ and ${\rm d}\tilde P = s_2 {\rm d}T_2$, we obtain
\begin{eqnarray} \label{Eq:lnTa}
\partial_i\Big( \ln(T_1 a) \Big) = 0, \hspace{1cm} \partial_i\Big( \ln(T_2 \tilde{a}) \Big) = 0.
\end{eqnarray}
Therefore, $T_1(\bold{x})a(\bold{x})$ and $T_2(\bold{x})\tilde{a}(\bold{x})$ should be constants when the subsystems are individually in thermal equilibrium. As discussed above, for global thermal equilibrium to exist, we should further have
\begin{equation}\label{Eq:GT}
    T_1(\bold{x})a(\bold{x}) = T_2(\bold{x})\tilde{a}(\bold{x}) = \mathcal{T}_0,
\end{equation}
where $\mathcal{T}_0$ (a constant) is the physical temperature of the full system. However, we note that we can solve the subsystem Ward identities more generally by requiring that the two subsystems have two different (constant) physical temperatures, i.e.
\begin{equation}\label{Eq:PE}
    T_1(\bold{x})a(\bold{x}) = T_a, \quad T_2(\bold{x})\tilde{a}(\bold{x}) = T_b.
\end{equation}

The effective metrics~\eqref{Eq:Effmet} should satisfy the coupling equations~\eqref{Eq:metriccoup} which take the form
\begin{align} \label{Eq:thermalcoup}
e^{-2\phi(\bold{x})}-a(\bold{x})^2 &=& \Bigg( \frac{e^{-\phi(\bold{x})} (d-1)\tilde{P}(T_2(\bold{x}))~ ~\gamma' }{\tilde{b}(\bold{x})^2} + \frac{(\gamma-\gamma') ~ e^{- 3 \phi(\bold{x})}  ~\et(T_2(\bold{x}))}{\tilde{a}(\bold{x})^2} \Bigg) \tilde{a}(\bold{x}) \tilde{b}(\bold{x})^{d-1}, \nonumber  \\
b(\bold{x})^2-1 &=& \Bigg( \frac{ (\gamma-(d-1)\gamma')~ e^{\phi(\bold{x})}~\tilde{P}(T_2(\bold{x}))}{\tilde{b}(\bold{x})^2} + \frac{ e^{-\phi(\bold{x})}~ \gamma' ~ \e(T_2(\bold{x}))}{\tilde{a}(\bold{x})^2} \Bigg) \tilde{a}(\bold{x}) \tilde{b}(\bold{x})^{d-1}, \nonumber \\
e^{-2\phi(\bold{x})}-\tilde{a}(\bold{x})^2 &=& \Bigg( \frac{e^{-\phi(\bold{x})} (d-1){P}(T_1(\bold{x})) ~\gamma' }{{b}(\bold{x})^2} + \frac{(\gamma-\gamma') e^{-3 \phi(\bold{x})} ~{\e}(T_1(\bold{x}))}{{a}(\bold{x})^2} \Bigg) {a}(\bold{x}) {b}(\bold{x})^{d-1}, \nonumber \\
\tilde{b}(\bold{x})^2-1 &=& \Bigg( \frac{ (\gamma-(d-1) \gamma') ~e^{\phi(\bold{x})} ~{P}(T_1(\bold{x}))}{{b}(\bold{x})^2} + \frac{e^{-\phi(\bold{x})}~ \gamma' ~ \e(T_1(\bold{x}))}{{a}(\bold{x})^2} \Bigg) {a}(\bold{x}) {b}(\bold{x})^{d-1}.
\end{align}
The above equations and~\eqref{Eq:GT} thus determine the six variables $a(\bold{x})$, $b(\bold{x})$, $\tilde{a}(\bold{x})$, $\tilde{b}(\bold{x})$, $T_1(\bold{x})$ and $T_2(\bold{x})$, and therefore the full thermal equilibrium which is labeled just by $\mathcal{T}_0$, the physical temperature. If we use~\eqref{Eq:PE} instead of~\eqref{Eq:GT}, we obtain a more general family of solutions labeled by two (constant) parameters, namely $T_a$ and $T_b$, the physical temperatures of the subsystems. In the following subsection, we will identify these with pseudo-equilibrium states.

The energy-momentum tensor of the full system can be readily obtained from~\eqref{Eq:fullem} and~\eqref{Eq:fullem1}. It is easy to check that it assumes the equilibrium form
\begin{eqnarray}\label{Eq:ThermalFullEMT}
T^\mn = \Big( e^{-2 \phi(\bold{x})} \etot(\bold{x}), \underbrace{\mathcal{P}(\bold{x}), \cdots, \mathcal{P}((\bold{x}))}_{d-1} \Big),
\end{eqnarray}
where $\etot$ and $\mathcal{P}$ can be readily identified with the local energy density and pressure of the full system. We recall that the democratic effective metric coupling ensures that the full energy-momentum tensor should be conserved in the physical background metric if the subsystem energy-momentum tensors are conserved in their respective effective metrics. Therefore, given that~\eqref{Eq:GT}, or more generally~\eqref{Eq:PE}, ensures that the subsystem Ward identities~\eqref{Eq:subsystemWI} are satisfied, the full energy-momentum tensor~\eqref{Eq:ThermalFullEMT} must be conserved in the physical background metric~\eqref{Eq:backgroundmetric}, which leads to
\begin{eqnarray}
\frac{\partial_i \mathcal{P}}{\etot+\mathcal{P}} - \partial_i \phi = 0.
\end{eqnarray}
When the subsystems have the same physical temperature $\mathcal{T}_0$, the complete solution and therefore the full energy-momentum tensor~\eqref{Eq:ThermalFullEMT} are labeled by $\mathcal{T}_0$. In this case, the thermodynamic identities~\eqref{Eq:thermoidentities-full} for the full system are well defined, and they imply that
\begin{eqnarray}
\frac{\partial_i \mathcal{T}}{\mathcal{T}} - \partial_i \phi =\partial_i({\rm ln}(\mathcal{T}e^{-\phi}))= 0, 
\end{eqnarray}
i.e. 
\begin{equation}\label{Eq:Tf}
    \mathcal{T(\bold{x})}e^{-\phi(\bold{x})} = T_f,
\end{equation}
a constant. \textit{To prove thermodynamic consistency, we need to show that if $T_a = T_b = \mathcal{T}_0$, then we must have $T_f =\mathcal{T}_0$.} Obviously, this ensures that the requirement of global thermal equilibrium~\eqref{Eq:TC} is satisfied. \textit{To prove statistical consistency, we need to show that if $T_a = T_b = \mathcal{T}_0$, then $\mathcal{S}$, the entropy density of the full system obtained from the thermodynamic identities~\eqref{Eq:thermoidentities-full} must satisfy~\eqref{Eq:SC}.}

Both thermodynamic and statistical consistencies follow from the general form of the full energy-momentum tensor. The latter given by~\eqref{Eq:fullem} and~\eqref{Eq:fullem1} with one contravariant and one covariant index indicates that the terms in $ \Delta K$ are always diagonal and appear with opposite signs for $T^0_{\,0}$ and $T^i_{\,i}$, i.e., for $\mathcal{E} $ and $ \mathcal{P}$. This leads to the identity
\begin{eqnarray}
(\e(\bold{x}) + P(\bold{x})) a(\bold{x}) b(\bold{x})^{d-1} + (\et (\bold{x})+ \tilde{P}(\bold{x})) \tilde{a(\bold{x})} \tilde{b}(\bold{x})^{d-1} = (\etot(\bold{x}) + \mathcal{P}(\bold{x})) e^{-\phi(\bold{x})}.
\end{eqnarray}
The thermodynamic identities of the subsystems and the full system then imply that
\begin{eqnarray}\label{Eq:TSfull}
T_1(\bold{x}) s_1(\bold{x}) a(\bold{x}) b(\bold{x})^{d-1} + T_2(\bold{x}) s_2(\bold{x})\tilde{a}(\bold{x})  \tilde{b}(\bold{x})^{d-1} =\mathcal{T}(\bold{x})\mathcal{S}(\bold{x}) e^{-\phi(\bold{x})}.
\end{eqnarray}
Assuming~\eqref{Eq:GT} and using~\eqref{Eq:Tf}, we thus obtain that
\begin{eqnarray}
\mathcal{T}_0(s_1(\bold{x}) b(\bold{x})^{d-1} + s_2(\bold{x}) \tilde{b}(\bold{x})^{d-1})= T_f\mathcal{S}(\bold{x}) .
\end{eqnarray}
Therefore, we should have
\begin{equation}
    T_f = \lambda \mathcal{T}_0, \quad \mathcal{S}(\bold{x}) =\lambda^{-1}\left( s_1(\bold{x}) b(\bold{x})^{d-1} + s_2(\bold{x}) \tilde{b}(\bold{x})^{d-1}\right),
\end{equation}
with $\lambda$, an arbitrary positive constant. Using~\eqref{Eq:Tf} again, we obtain that 
\begin{equation}
    \mathcal{T}(\bold{x}) = \lambda e^{\phi(\bold{x})}\mathcal{T}_0, \quad \mathcal{S}(\bold{x}) =\lambda^{-1}\left( s_1(\bold{x}) b(\bold{x})^{d-1} + s_2(\bold{x}) \tilde{b}(\bold{x})^{d-1}\right),
\end{equation}
which means that the local temperature and entropy densities of the full system are determined up to multiplication by a constant $\lambda$ and its inverse, respectively. However, this multiplicative ambiguity is a fundamental feature of the definition of temperature and the entropy density, and is present even for the subsystems, as should be clear from the thermodynamic identities~\eqref{Eq:thermoidentities-full} and~\eqref{Eq:thermoidentities}. This ambiguity can only be fixed by the choice of the temperature scale (which is independent of the couplings). Here, $\lambda$ can easily be fixed by taking the decoupling limit $\gamma,\gamma' \rightarrow 0$ where it is obvious that $\lambda = 1$, as the full energy momentum tensor is simply the sum of those of the subsystems. Therefore, we must have
\begin{equation}
    \mathcal{T}(\bold{x})e^{-\phi(\bold{x})} = \mathcal{T}_0, \quad \mathcal{S}(\bold{x}) =s_1(\bold{x}) b(\bold{x})^{d-1} + s_2(\bold{x}) \tilde{b}(\bold{x})^{d-1},
\end{equation}
implying that both the thermodynamic consistency condition~\eqref{Eq:TC} and the statistical consistency condition~\eqref{Eq:SC} are satisfied when the global equilibrium condition~\eqref{Eq:GT} is valid.

\subsection{The persistence of pseudo-equilibrium}

Although the democratic effective coupling scheme leads to a thermal equilibrium which is consistent with the principles of thermodynamics and the statistical definition of entropy, we obtain a family of pseudo-equilibrium solutions as well, which are labeled by $T_a$ and $T_b$, the temperatures of the subsystems. These represent \textit{pseudo-equilibrium states}. In such general pseudo-equilibrium states, we can still define the total entropy density as 
\begin{eqnarray}
\mathcal{S}(\bold{x}) = s_1(\bold{x}) b(\bold{x})^{d-1} + s_2(\bold{x}) \tilde{b}(\bold{x})^{d-1}.
\end{eqnarray}
Since the individual subsystems are in thermal equilibrium, the total entropy can be defined as the sum of the subsystem entropies once the effective volume factors are included (recall that from the point of view of the effective metrics, the subsystems are isolated). For full statistical consistency, it is necessary to show that \textit{of all pseudo-equilibrium states with fixed energy, the global equilibrium state satisfying~\eqref{Eq:TC} should have the maximum entropy.} If the latter is true, then it ensures that the microcanonical ensemble is well-defined. In the following section, we will prove that this consistency requirement is also satisfied when two subsystems with \textit{arbitrary} equations of state couple via the democratic effective metric coupling. 

However, there is another problem of dynamical origin. More generally, if the two subsystems have entropy currents $s_1^{\mu}$ and $s_2^{\mu}$  (which in equilibrium take the forms $s_1^{\mu} = s_1 u^{\mu}$ and $s_2^{\mu} = s_2 \tilde{u}^{\mu}$) that satisfy $\nabla_{\mu} s_1^{\mu} \geq 0$ and  $\tilde{\nabla}_{\mu} s_2^{\mu} \geq 0$ in their respective backgrounds, then it is easy to see that the total entropy current $\mathcal{S}^{\mu}$ can be defined as
\begin{eqnarray} \label{Eq:ent-current}
    \mathcal{S}^{\mu} := \frac{\sqrt{- g}}{\sqrt{- g^{(B)}}} s_1^{\mu} + \frac{\sqrt{-\tilde{g}}}{\sqrt{-g^{(B)}}} s_2^{\mu},
\end{eqnarray}
and it satisfies
\begin{equation}
    \nabla^{(B)}_{\mu} \mathcal{S}^{\mu} \geq 0.
\end{equation}
In equilibrium, $\mathcal{S}^{\mu} = \mathcal{S} U^{\mu}$ with $U^\mu$ being the thermal frame of the full system. Note that the definition of entropy current~\eqref{Eq:ent-current} holds generally even out of equilibrium. It is useful to recall the discussion at the end of the previous section in which we emphasized that the formulation gives a non-fluctuating effective metric only in the large $N$ limit. Away from the large $N$ limit, the two effective background metrics fluctuate, although the physical background metric is fixed by definition. So, away from the large $N$ limit, only the total entropy current and the total conserved currents are meaningful, and not the individual subsystem entropy currents and conserved currents. 

Usually, the presence of an entropy current ensures thermalization as entropy increases dynamically, and the global equilibrium has maximum entropy. The issue in the democratic coupling scheme is that in the large $N$ limit, where the effective background metrics cannot fluctuate, each subsystem has its own entropy current. Therefore, each subsystem can thermalize individually, and the full system can reach a pseudo-equilibrium state without attaining global equilibrium where the physical temperatures of the subsystems are not equal, although the global equilibrium has maximum entropy. Therefore, in the large $N$ limit, the full system can be stuck in a pseudo-equilibrium state without thermalizing. This issue can be resolved by first noting that the large $N$ limit implicitly implies large energy. We will show in Sec.~\ref{Sec:DT}, that a typical non-equilibrium state does thermalize effectively in the large $N$ limit.

\section{Global equilibrium maximizes the total entropy for the isolated hybrid system}\label{Sec:GEM}

In the preceding section, we have shown that global equilibrium satisfies both thermodynamic as well as statistical consistencies. However, the interaction of the individual sectors via their effective metric allows the subsystems to equilibrate to a pseudo-equilibrium state where they have different physical temperatures $ \temp$ and $ \tempt$. This makes it less obvious that at global equilibrium, the total entropy of the full system is maximum. 
For our coupling scheme to be physically viable, one needs to ensure that the global equilibrium must have maximum entropy in a microcanonical ensemble with constant total energy. In what follows, we prove that this is indeed the case.  

In a microcanonical ensemble, the isolation of the full system implies that the total energy $E_0$ is a constant while the individual sectors interact via their effective metric and equilibrate with temperatures $ \temp$ and $\tempt$, leading the full system to a pseudo-equilibrium state. The condition that the total energy is constant relates the temperatures $\temp$ and $\tempt$ of the two subsectors. Hence, in a microcanonical ensemble, the space of pseudo-equilibrium solutions is characterized by one parameter, say, $\lambda$. We will prove via extremization of the total entropy on the manifold of pseudo-equilibrium solutions with fixed energy $E_0$ (and parameterized by $ \lambda$) that the global equilibrium state, where both subsystems have the same physical temperature, is the unique maximum entropy state in a microcanonical ensemble.

We demonstrate this for an isolated system $ \mathcal{F}$ with two subsectors $ \mathcal{U}$ and  $\tilde{\mathcal{U}}$ defined by arbitrary equations of states, $P(\e)$ \text{and} $\pt(\et)$, see \Cref{fig:micro-picture}. The physical background metric for the full system is the Minkowski metric
\begin{eqnarray}
g_\mn^{(B)} = \eta_\mn = \text{diag}(-1, \underbrace{1, \cdots, 1}_{d-1}).
\end{eqnarray}
In the pseudo-equilibrium states with a fixed total energy, the two sectors live in their individual effective metrics $ g_\mn $ and $\tilde{g}_\mn$ given by 
\begin{align}\label{eff-metrics-lambda}
    g_{\mn} = \text{diag} \left( -a(\lambda)^2, \underbrace{b(\lambda)^2, \cdots, b(\lambda)^2}_{d-1} \right), \hspace{1cm}
    \gt_{\mn} = \text{diag} \left( -\At(\lambda)^2, \underbrace{\Bt(\lambda)^2, \cdots, \Bt(\lambda)^2}_{d-1}\right).
\end{align}
(Recall the discussion in the previous section.) Above, $\lambda$ parametrizes the space of pseudo-equilibrium states. The two sectors interact via the metric coupling \eqref{Eq:thermalcoup} and generically equilibrate to a pseudo-equilibrium state in which the two subsystems have effective temperatures $ T_1(\lambda) $ and $ T_2(\lambda)$, respectively. The physical temperatures of these subsystems are $\temp = a T_1 $ and $\tempt = \tilde{a} T_2 $, respectively. 

The goal is to find the value of $\lambda^*$, the value of $\lambda$ at which the total entropy density 
\begin{equation}
    \mathcal{S}(\lambda) = s_1(T_1(\lambda))b(\lambda)^{d-1}+ s_2(T_2(\lambda))\tilde{b}(\lambda)^{d-1},
\end{equation}
is maximized subject to the constraint that the total energy density $\mathcal{E}$ given by
\begin{eqnarray} \label{Eq:fullenepresure}
\mathcal{E} &=& \epsilon(T_1(\lambda)) a(\lambda) b(\lambda)^{d-1} + \tilde{\epsilon}(T_2(\lambda)) \tilde{a(\lambda)} \tilde{b(\lambda)}^{d-1} \nonumber \\ &+& \frac{\gamma}{2} \Bigg( \frac{\epsilon(T_1(\lambda)) \tilde{\epsilon}(T_2(\lambda))}{a(\lambda)^2 \tilde{a}(\lambda)^2} + \frac{(d-1) P(T_1(\lambda)) \tilde{P}(T_2(\lambda))}{b(\lambda)^2 \tilde{b(\lambda)^2}}  \Bigg) a(\lambda) b(\lambda)^{d-1}\tilde{a}(\lambda)  \tilde{b}(\lambda)^{d-1} \nonumber \\
&+& \frac{\gamma'}{2} \Bigg(  \frac{(d-1) P(T_1(\lambda))}{b(\lambda)^2}  -\frac{\epsilon(T_1(\lambda))}{a(\lambda)^2}\Bigg) \Bigg(  \frac{(d-1)\tilde{P}(T_2(\lambda))}{\tilde{b}(\lambda)^2} -\frac{\tilde{\epsilon}(T_2(\lambda))}{\tilde{a}(\lambda)^2} \Bigg)\nonumber\\&&\qquad \times a(\lambda) b(\lambda)^{d-1} \tilde{a}(\lambda)  \tilde{b}(\lambda)^{d-1}.
\end{eqnarray}
is a constant\footnote{Since we are considering homogeneous configurations, the total energy density should be held fixed and the total entropy density should be maximized. It will be clear from the proof that it is sufficient to prove the maximization of entropy in global thermal equilibrium in homogeneous equilibrium as the proof can be implemented locally.}. We recall that $a(\lambda)$, $b(\lambda)$, $\tilde{a}(\lambda)$ and $\tilde{b}(\lambda)$ should be obtained by solving the coupling equations~\cref{Eq:thermalcoup} (with $\phi =0$) which are fully specified by the subsystem temperatures, $T_1(\lambda)$ and $T_2(\lambda)$.

An elegant way to find $\lambda_*$ is as follows. At $\lambda = \lambda_*$, we have
\begin{equation}\label{Eq:SEmax}
    \frac{{\rm d}\mathcal{S}}{{\rm d}\lambda} = 0, \quad \frac{{\rm d}\mathcal{E}}{{\rm d}\lambda} = 0,
\end{equation}
as $\mathcal{S}$ is maximum at $\lambda = \lambda_*$ and $\mathcal{E}$ is independent of $\lambda$. Furthermore, as $a(\lambda)$, $b(\lambda)$, $\tilde{a}(\lambda)$ and $\tilde{b}(\lambda)$ satisfy the four coupling equations for all values of $\lambda$, the derivatives of these coupling equations with respect to $\lambda$ also should vanish at any $\lambda$, and hence at $\lambda =\lambda_*$ also. From \eqref{Eq:SEmax} and the derivatives of the four coupling equations, we obtain the following system of \textit{homogeneous linear} equations which take the form
\begin{equation}\label{Eq:Lin-Hom}
    \mathcal{Q}(\lambda)\cdot\mathcal{A} = 0
\end{equation}
where $\mathcal{Q}$ is a $6\times6$ matrix and $\mathcal{A}$ is the vector given by
\begin{equation}
    \mathcal{A}  = \left\{ \frac{{\rm d} a}{{\rm d}\lambda},\frac{{\rm d} b}{{\rm d}\lambda}, \frac{{\rm d} \tilde{a}}{{\rm d}\lambda},\frac{{\rm d} \tilde{b}}{{\rm d}\lambda}, \frac{{\rm d} T_1}{{\rm d}\lambda},\frac{{\rm d} T_2}{{\rm d}\lambda}\right\}.
\end{equation}
In order to obtain the above equation we need to convert derivatives of the subsystem energies and pressures to derivatives of $T_1$ and $T_2$ which can be achieved by using the thermodynamic identities. Explicitly, the derivatives of the subsystem energies and pressures are
\begin{align}
    &\frac{d \e}{d \lambda} = \frac{1}{\cs(\lambda)^2} \frac{d P}{d \lambda}, 
    &&\frac{d \et}{d \lambda} = \frac{1}{\cst(\lambda)^2} \frac{d \pt}{d \lambda},\nonumber\\&
    \frac{d P}{d \lambda} = \frac{P(\lambda) + \e(\lambda)}{T_1(\lambda)}  \frac{d T_1}{d \lambda}, 
    &&\frac{d \pt}{d \lambda} = \frac{\pt(\lambda) + \et(\lambda)}{T_2(\lambda)}  \frac{d T_2}{d \lambda}.
\end{align}
where the speed of sound of the subsectors are $c_s^2 = \frac{{ d}P}{{ d}\e}$ and $\cst^2=\frac{d \pt}{d \et}$. 

Finally, we note that at $\lambda =\lambda_*$, the set of homogeneous linear equations \eqref{Eq:Lin-Hom} has a solution at $\lambda =\lambda_*$ iff
\begin{equation}
    {\rm det} \mathcal{Q}(\lambda) =0.
\end{equation}
Note that the above equation is an equation for only one variable $\lambda$. 
Explicitly, we find that
\begin{align}\label{det=0}
    \text{det} \mathcal{Q} =& \Big(a(\lambda )
   T_1(\lambda )-\At(\lambda ) T_2(\lambda )\Big)\mathcal{X} = \left(T_a(\lambda)-T_b(\lambda)\right)\mathcal{X}
\end{align}
where the expression for $\mathcal{X}$ is lengthy and can be found in Appendix \ref{Sec:Lengthy}. We readily see that if the global equilibrium condition $T_a =T_b$ is satisfied, then $\det\mathcal{Q}$ vanishes. This should correspond to a maximum of the total entropy if the second derivative of the total entropy with respect to $\lambda$ is negative at $\lambda_*.$

Given that the full system satisfies thermodynamic consistency, the global equilibrium condition can always be satisfied for a certain value of $\lambda$ in the manifold of pseudo-equilibrium states with a fixed total energy, as shown in the previous section. We also find that for most equations of states of the subsystems yielding positive specific heats, the total entropy $\mathcal{S}(\lambda)$ is a concave function of $\lambda$ if, e.g.~$\lambda$ is chosen to be one of the subsystem temperatures. Therefore, $\mathcal{S}(\lambda)$ can have only one maximum, which corresponds to the global equilibrium. This completes the proof that the microcanonical ensemble is indeed consistent when two subsystems are coupled via the democratic effective metric coupling, as the global equilibrium state is the unique maximum entropy state.

\begin{figure}[tbp]
         \includegraphics[width=0.49\linewidth]{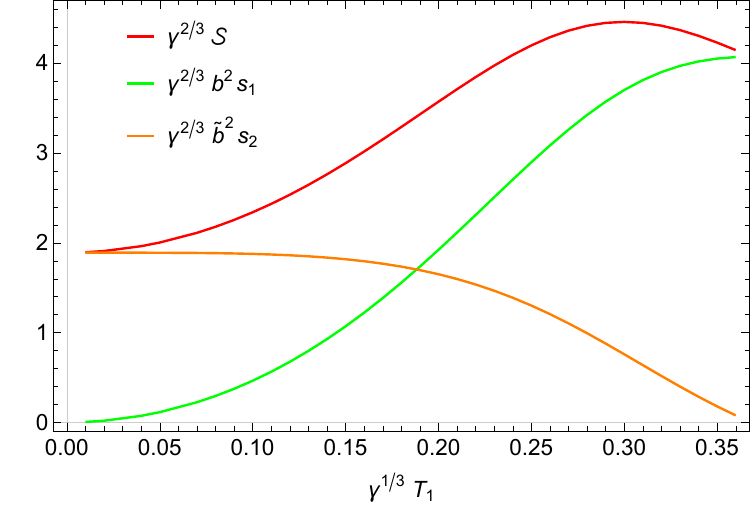}
         \includegraphics[width=0.49\linewidth]{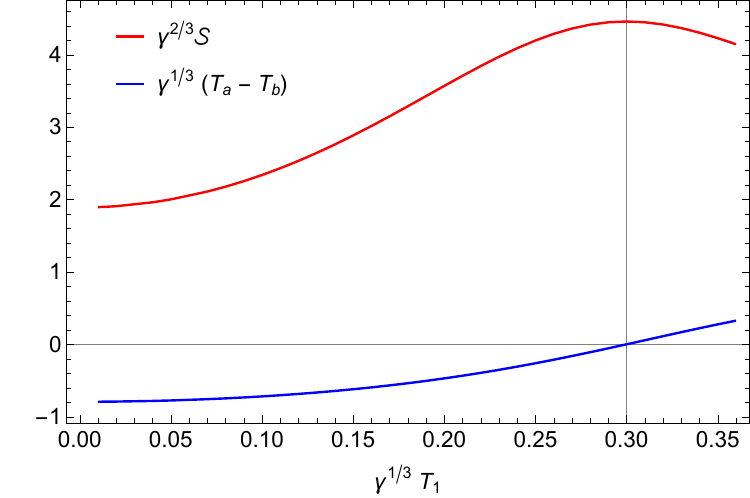}
     \caption{Pseudo-equilibrium states for two 3-dimensional conformal subsystems coupled via the democratic effective metric coupling. Explicitly, $\e = 2 \p = n_1 T_1^3$, $\et = 2 \pt = n_2 T_2^3$ and we set $\gamma'=2$, $\gamma=1$, $n_1=10$, $n_2=1$.  Left: subsystem entropies and the full system entropy as functions of $T_1$. Right: full system entropy and global equilibrium measure ($\A T_1- \At T_2 = T_a - T_b$) as functions of $T_1$. Note that the total entropy maximizes precisely when the global equilibrium condition is satisfied.
     }
     \label{fig: n1-neq-n2}
\end{figure}

For illustration, we can consider the case of two subsystems with conformal equations of states $\epsilon = n_1 T_1^3$ and $\tilde\epsilon = n_2 T_2^3$ in $2+1-$dimensions. We choose $n_1 = 10$ and $n_2 =1$. In the left panel in \Cref{fig: n1-neq-n2}, we note that while one of the subsystem entropies monotonically increases with its effective temperature, that of the other decreases as a consequence of the fixed total energy. With the total energy $\mathcal{E}$ kept constant and setting $\lambda = T_1$, the total entropy $\mathcal{S}$ has been plotted alongside $T_a - T_b$ as a function of $T_1$ in the right panel of \Cref{fig: n1-neq-n2}. We observe that $T_a - T_b$ indeed vanishes exactly where $\mathcal{S}$ is maximum. We also note that the total entropy is a concave function and has only one maximum corresponding to the global equilibrium.

As a passing remark, we point out that the local temperatures of the subsystems, namely $T_1$ and $T_2$ are bounded in order for the metric coupling equations to have physical solutions  \cite{Kurkela:2018dku}. However, the physical temperatures $\mathcal{T}_0$ of the full system in global thermal equilibrium can be arbitrarily large, and $\mathcal{T}_a$ and $\mathcal{T}_b$ individually in pseudo-equilibrium states as well. Of course, $T_1$ and $T_2$ are also bounded if the full energy density of the system is specified. 

It is also worth emphasizing that the full system has an emergent conformality at large $\gamma\mathcal{E}_0$ which is borne out by the full equation of state \cite{Kurkela:2018dku} and also by the hydrodynamic and quasinormal mode dispersion relations \cite{Kurkela:2018dku, Mondkar:2021qsf}. This emergent conformality at large $\gamma\mathcal{E}_0$ is observed for arbitrary equations of states of the subsystems.

\section{Dynamics and thermalization}\label{Sec:DT}
We have demonstrated that the democratic effective metric coupling satisfies both thermodynamic and statistical consistencies, implying that when the coupled subsystems have the same physical temperature, then the thermodynamic identities of the full system are satisfied, with the total entropy being consistent with its interpretation in statistical mechanics as the logarithm of the number of microstates. Furthermore, although the subsystems can thermalize individually to different physical temperatures, giving rise to pseudo-equilibrium states in the large $N$ limit, the global equilibrium has the maximum entropy when the full system is isolated. Nevertheless, it is necessary to study whether the full system can \textit{generically} reach the global equilibrium in the large $N$ limit.

In the present section, we will show that in the large $N$ limit, a typical non-equilibrium state of the isolated hybrid system thermalizes to global equilibrium. In order to show this, we consider the isolated hybrid system with total energy $E_0$, which remains constant during time evolution. Let the initial non-equilibrium state of the full system be generic, and its coarse-grained total entropy be $S_{in}$. After entropy production ceases, this state reaches a homogeneous pseudo-equilibrium state with total entropy density $S_f$. We denote $S_{geq}$ as the total entropy density if the full system is in global equilibrium at the same total energy. We will show that for a typical initial non-equilibrium state
\begin{equation}\label{Eq:Sprod}   \lim_{\mathcal{E}_0\rightarrow\infty}\frac{S_f- S_{in}}{S_{geq}- S_{in}} = 1,
\end{equation}
i.e.~the final pseudo-equilibrium state coincides with the global equilibrium state in the limit in which the fixed total average energy density $\mathcal{E}_0$ is taken to infinity for a typical initial isolated non-equilibrium state. Given that the global thermal equilibrium state is the unique maximum entropy state at any fixed total energy, the final state is homogeneous and that the total energy is conserved in the evolution of the full isolated system, the above implies that the final state coincides with the global thermal equilibrium state in the limit in which the initial average total energy density tends to infinity.

We will illustrate \eqref{Eq:Sprod} with \textit{typical homogeneous} initial non-equilibrium states. In this case, the total energy density $\mathcal{E}_0$ remains constant with time evolution since it preserves homogeneity. Note that the average energy and entropy densities should be large in a typical state in the large $N$ limit. Therefore, \eqref{Eq:Sprod} is tantamount to thermalization in the large $N$ limit for a typical initial state.

\subsection{Hybrid BRSSS setup}

The simplest setup to study hybrid thermalization in the large $N$ limit involves the causal coarse-grained MIS description for each subsystem. This setup has been previously utilized to study the hybrid hydrodynamic attractor and hydrodynamization \cite{Mitra:2020mei,Mitra:2022uhv}. When the subsystems are conformal, the BRSSS formulation \cite{Baier:2007ix}, which is a modified version of the MIS theory, allows for a coarse-grained entropy.\footnote{For simplicity, we will use a truncated version of BRSSS theory \cite{Baier:2007ix} with $\kappa$ and $\lambda_1$ set to zero. We retain $\tau_\pi$. The other second order transport coefficients are irrelevant in our study as we do not have vorticity.} We study the particular case of two three-dimensional conformal subsystems, each described by BRSSS theory and coupled via the democratic effective metric coupling. We also focus on the process of homogeneous thermal relaxation.

The full three-dimensional system lives in the flat Minkowski metric,
\begin{equation}
    g_{\mu\nu}^{(B)} \equiv \eta_\mn = \text{diag}(-1,1,1),
\end{equation}
which is the physical metric. Assuming homogeneity, the effective metrics of the subsystems take the form
 \begin{equation}
g_{\mn} = 
    \begin{pmatrix}
-a_{11}(t)^2 & 0 & 0\\
0 &  a_{22}(t)^2 & a_{23}(t) \\
0 & a_{23}(t) & a_{33}(t)^2
\end{pmatrix}
\hspace{0.5cm} \text{and} \hspace{0.5cm}
\gt_{\mn} = 
    \begin{pmatrix}
-\At_{11}(t)^2 & 0 & 0\\
0 &  \At_{22}(t)^2 & \At_{23}(t) \\
0 & \At_{23}(t) & \At_{33}(t)^2
\end{pmatrix}.
\end{equation}
We will see that we can consistently set the $g_{ti}$ and $\gt_{ti}$ components of the effective metrics to zero.

The energy-momentum tensors of the subsystems take the form
\begin{align} \label{Eq:diagstresstensor}
    t^{\mn} =  (\e + P)\uu^{\mu} \uu^{\nu} + P g^{\mn} + \py^{\mn} \hspace{0.6cm} \text{and} \hspace{0.6cm}
    \Tt^{\mn} =  (\et + \pt)\uut^{\mu} \uut^{\nu} + \pt \gt^{\mn} + \pyt^{\mn}.
\end{align}
Due to conformality, $\e = 2P$ and $\et = 2\pt$. Assuming homogeneity,\footnote{Note that homogeneity allows only relaxation processes, as hydrodynamic shear and sound modes require inhomogeneity.} the time-like fluid velocity $u^\mu(\Tilde{u}^\mu)$ of each subsystem 
takes the form
  \begin{eqnarray}
      u^\mu = \text{diag}(1/a_{11}(t),0,0 ) \hspace{0.5cm} \text{and} \hspace{0.5cm} 
       \Tilde{u}^\mu = \text{diag}(1/\Tilde{a}_{11}(t),0,0),
  \end{eqnarray}
and the non-equilibrium shear stress tensor capturing terms responsible for dissipation in \eqref{Eq:diagstresstensor} can be taken to be
       \begin{eqnarray}
\py^{\mu \alpha}g_{\alpha\nu} \equiv \py^{\mu}_{\nu} = 
    \begin{pmatrix}
0 & 0 & 0\\
0 &  - \frac{\py_d (t)}{2} & \py_{od} (t) \\
0 & \frac{\py_d (t)}{a_{33}(t)} + \frac{a_{22}(t)^2 \py_{od} (t)}{a_{33}(t)^2}&    \frac{\py_d (t)}{2}
\end{pmatrix}, \\
\pyt^{\mu \alpha}\gt_{\alpha\nu} \equiv \pyt^{\mu}_{\nu} = 
    \begin{pmatrix}
0& 0 & 0\\
0 & - \frac{\pyt_d (t)}{2} & \pyt_{od} (t) \\
0 & \frac{\pyt_d (t)}{\At_{33}(t)} + \frac{\At_{22}(t)^2 \pyt_{od} (t)}{\At_{33}(t)^2}&    \frac{\pyt_d (t)}{2}
\end{pmatrix}.
\end{eqnarray}
Above $\py_d$ and $\pyt_d$ denote the diagonal anisotropic pressure, whereas $\py_{od}$ and $\pyt_{od}$ denote the off-diagonal terms.  Note that $\py^{\mu}_{\nu}$ and $\pyt^{\mu}_{\nu}$ should be traceless as we assume conformality of both subsystems. Also, the Landau frame identification of $u^\mu$ and $(\Tilde{u}^\mu)$ as the velocities of the energy currents of the subsystems implies that these are transverse to 
{$\py^{\mu}_{\nu}$ and $\pyt^{\mu}_{\nu}$}, respectively. This also implies that we can set $g_{ti}$ and $\gt_{ti}$ components of the effective metrics to zero consistently in the coupling equations, as mentioned before.

The BRSSS equations of motion describing the relaxation of the shear stress tensors in the respective effective metrics are

\begin{eqnarray} 
 \Big[\pi^{\mn} + \tau_\pi D \pi^{\mn}  \Big]  + \frac{3}{2} \tau_\pi \pi^{\mn} \nabla_{\alpha}  u^{\alpha}  = -\eta \sigma^{\mn}, \label{Eq:chap6mis}\\
  \Big[\tilde{\pi}^{\mn} + \tilde{\tau}_\pi \tilde{D} \tilde{\pi}^{\mn}  \Big]  + \frac{3}{2} \tilde{\tau}_\pi \tilde{\pi}^{\mn} \tilde{\nabla}_{\alpha}  \tilde{u}^{\alpha}  = -\tilde{\eta} \tilde{\sigma}^{\mn} , \label{Eq:chap6mis-2}
\end{eqnarray}
where $\tau_\pi(\tilde{\tau}_\pi)$ and $ \eta(\tilde{\eta})$ are the relaxation time and shear viscosity, respectively. The shear tensors, 
$ \sigma^\mn$ and $ \tilde{\sigma}^\mn$, are defined as
\begin{eqnarray}
\sigma^\mn := \frac{1}{2}(\nabla_\alpha u_\beta + \nabla_\beta u_\alpha) \Delta^{\alpha \beta} \Delta^\mn - \frac{1}{2} \nabla_\lambda u^\lambda \Delta^\mn, \nonumber \\
\tilde{\sigma}^\mn := \frac{1}{2}(\tilde{\nabla}_\alpha \tilde{u}_\beta + \tilde{\nabla}_\beta \tilde{u}_\alpha) \tilde{\Delta}^{\alpha \beta} \tilde{\Delta}^\mn - \frac{1}{2} \tilde{\nabla}_\lambda \tilde{u}^\lambda \tilde{\Delta}^\mn \nonumber,
\end{eqnarray}
in the respective background metrics, and $\Delta^\mn = g^\mn + u^\mu u^\nu$ and $\tilde{\Delta}^\mn=\tilde{g}^\mn + \tilde{u}^\mu \tilde{u}^\nu$ are the spatial projections orthogonal to $u^\mu$ and $\tilde{u}^\mu$, respectively, defined in the respective effective metrics.

In the homogeneous thermal relaxation process in $2+1$-dimensions, the full system is governed by fourteen equations; namely, one Ward identity for each subsystem (the conservation of the energy-momentum tensor gives only one non-vanishing equation for each subsystem), two non-vanishing equations {from ~\eqref{Eq:chap6mis}, and~\eqref{Eq:chap6mis-2}} for each subsystem giving the relaxation of the shear stress tensor and the {eight} non-vanishing coupling equations~\eqref{Eq:metriccoup}. These equations are given in \Cref{Sec:Lengthy}. Utilizing these equations, we can solve for the fourteen independent variables, namely the seven variables $\e$, $\py_{d}$, $\py_{od}$, $a_{11}$, $a_{22}$, $a_{33}$ and $a_{23}$ of the first subsystem and the seven corresponding ones for the other. To initialize, we need to specify only the initial values of six variables, namely $\e$, $\py_d$, and $\py_{od}$ of the first subsystem and the three corresponding ones for the other subsystem.  This is because the variables of the effective metrics are obtained via the algebraic coupling equations.

The subsystem Ward identities ensure the conservation of the full energy-momentum tensor in the physical background metric, the Minkowski space, as discussed before. The latter simply implies that the full (homogeneous) energy density is a constant.

The entropy densities $ s_1$, $ s_2$, and $ \mathcal{S}$ of the subsystems and the full system are as follows:
\begin{eqnarray}\label{Eq:EntropyRelation}
s_1 = \frac{\e + P}{T} - \frac{\tau_\pi}{4 \eta T} \pi^\mn \pi_\mn, \\
s_2 = \frac{\et + \tilde{P}}{\tilde{T}} - \frac{\tilde{\tau}_\pi}{4 \tilde{\eta} \tilde{T}} \tilde{\pi}^\mn \tilde{\pi}_\mn, \\
\mathcal{S} = \frac{\sqrt{-g}}{\sqrt{-g^{(B)}}} s_1 + \frac{\sqrt{-\tilde{g}}}{\sqrt{-g^{(B)}}} s_2.
\end{eqnarray}
In homogeneous relaxation, the above quantities 
grow with time.

For numerical simulations in $2+1$-dimensions, we choose parameters that are lead to the first subsystem being strongly self-interacting and the second one being weakly self-interacting. Both subsystems are conformal. We choose the equations of states of the two subsystems so that $\e/T^3 = n_1 = \et/\tilde{T}^3 = n_2 =2$.  The shear viscosity and the relaxation times of the subsystems are chosen as shown in Table~\ref{tab:Tab2}. We proceed to numerically solve the algebraic-differential system (explicitly given in~\cref{Sec:Lengthy}), such that the total energy density is conserved to a precision of $10^{-6}$.

\begin{table}[h!]
  \centering
  \begin{tabular}{|c||c|}
    \hline
    Subsystem-1 & Subsystem-2 \\
    \hline 
    $\eta =\frac{1}{4 \pi} \frac{\e + P}{T} $ &  $\tilde{\eta} =\frac{100}{4\pi} \frac{\et + \tilde{P}}{\tilde{T}} $ \\  [2mm]
    $\tau_\pi = \frac{2-\ln 2}{2 \pi T} $ & $\tilde{\tau}_\pi = \frac{500}{4 \pi \tilde{T}}$ \\   [2mm]
    $n_1 = 2$ & $n_2 = 2$ \\
    \hline 
  \end{tabular}
  \caption{The subsystem parameters used in numerical simulations.}
  \label{tab:Tab2}
\end{table}

\begin{figure}[tbp]
     \includegraphics[width=0.5\textwidth]{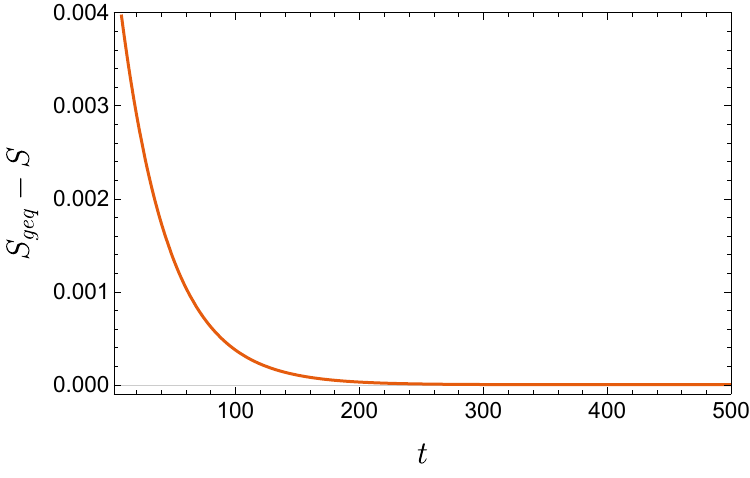}
 \includegraphics[width=0.48\textwidth]{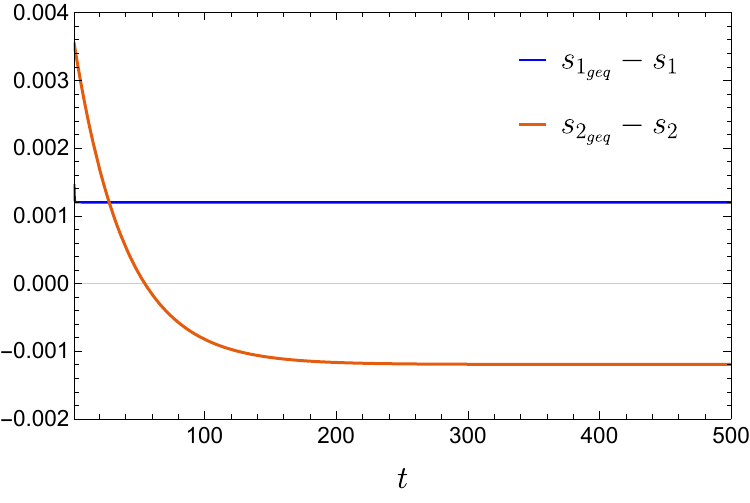}
    \caption{Evolution of the
    full system entropy density $\mathcal{S}$ (left) and the subsystem entropy densities $s_1$ and $s_2$ (right) for fixed energy density $\mathcal{E}_0 = 0.991$ and anisotropy $A1$ given in \Cref{tab:Tab1}. \TM{The inset plot in the left figure shows the evolution of the subsystem entropy density at early time.} Here we have fixed $\gamma = 1$ and $\tilde{\gamma} = -2$. We note that although the full entropy density reaches very close to the global equilibrium value, the subsystem entropy densities are away from the global equilibrium value. The final state is a pseudo-equilibrium state. We can explicitly check that the final state is typically much closer to the global equilibrium state for larger values of $\mathcal{E}_0$.}
     \label{Fig:EvolutionPlot}
 \end{figure}

\subsection{Thermalization for large energy densities}

We want to study the dynamics of homogeneous thermal relaxation for typical initial conditions for the full isolated system. The full energy density $\mathcal{E}_0$ during this process is fixed (in the physical Minkowski space) as discussed before. Recall that we need to specify the six variables, namely $\e$, $\py_d$, and $\py_{od}$ of the first subsystem and the three corresponding ones for the other subsystem. In order to choose typical initial conditions, we can proceed as follows.
\begin{itemize}
    \item Choose the diagonal and off-diagonal components of the shear-stress tensors, namely $\py_d$ and $\py_{od}$ of the first subsystem and the two corresponding ones of the second subsystem randomly.
    \item Choose a value of $\mathcal{E}_0$, the total energy density.
    \item Choose a value of $\delta := (a_{11} T - \tilde{a}_{11} \tilde{T})/(a_{11} T + \tilde{a}_{11} \tilde{T})$, where we note that $-1\leq \delta \leq 1$.
\end{itemize}
The above gives six conditions for specifying the six necessary initial conditions. 

We are interested in studying the final state obtained at the end of the relaxation process at large values of $\mathcal{E}_0$, the total energy density. Therefore, for a fixed (random) choice of the four variables of the subsystem stress tensors and $\delta$, we should study the characteristics of the final pseudo-equilibrium state as we increase $\mathcal{E}_0$. For illustrative purposes, we set $\delta =0$. We emphasize that our results are valid for any value of $\delta$ characterizing the initial conditions~\footnote{The choice of $\delta =0$ makes sense to study typical evolutions as the initial conditions can be readily interpreted as a perturbation about the global equilibrium state, the unique maximum entropy state.}.

We illustrate our results with the choices of the variables of the subsystem stress tensors shown in Table \ref{tab:Tab1}. In what follows, we demonstrate results for these choices of the four variables and $\delta =0$. Our results do hold for arbitrary choices of these variables. These variables should be within bounds (that depend on $\gamma\mathcal{E}_0$) so that the metric coupling equations have physical solutions.\footnote{Note that the shear stress tensor of the full system is unbounded, although the subsystem stress tensors in the respective effective metrics are bounded. This is exactly similar to the case of global thermal equilibrium, where the effective temperatures of the subsystems are bounded, although the physical temperatures of the full system (and the subsystems) are not.} Also note that increasing $\mathcal{E}_0$ at fixed $\gamma$ is the same as increasing $\gamma$ at fixed $\mathcal{E}_0$, as $\gamma$ is the only dimensionful parameter of the theory (we set $\gamma' = 2 \gamma$).

\begin{table}[h!]
  \centering
  \begin{tabular}{|c|c|c|c|c|c|}
    \hline
   &Color code in figure &$\pi_d$ & $\pi_{od}$ & $\tilde{\pi}_{d}$ & $\tilde{\pi}_{od}$ \\
    \hline
  A1&  Red & $0.1$ & $0.1$ & $0.01$ & $0.01$ \\
  A2&  Blue & $0.2$ & $0.01$ & $0.04$ & $0.005$ \\
    A3&   Orange & $0.03$ & $0.18$ & $0.004$ & $0.01$ \\
    \hline
  \end{tabular}
  \caption{Different anisotropies introduced in each of the subsystems.}
  \label{tab:Tab1}
\end{table}

\begin{figure}[H]
 \centering
    \includegraphics[width=0.5\textwidth]{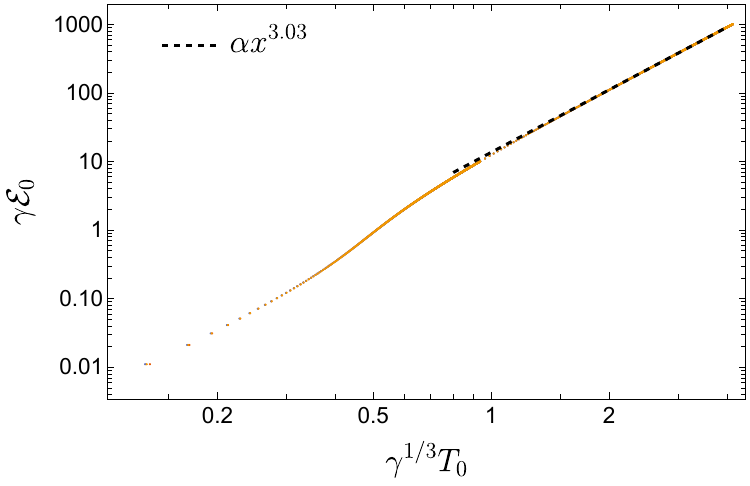}
    \caption{Total energy density as a function of $T_0$ for typical initial state with $\delta = 0$ for the three cases. For the three different anisotropies in \Cref{tab:Tab2}, the curves overlap.}
     \label{Fig:TAnisoFE}
 \end{figure}

Before studying the final state obtained after thermal relaxation, it is useful to see the statistical characteristics of the initial state. Enforcing $\delta =0$ for the initial state, we obtain $a_{11} T = \tilde{a}_{11} \tilde{T} := T_0$. Note that as the initial state is a non-equilibrium state, $T_0$ is not the physical temperature. At large $\mathcal{E}_0$ (or equivalently at large $\gamma$), the full system attains emergent conformality at global thermal equilibrium \cite{Kurkela:2018dku}. A priori, it is not expected that $T_0 \sim \mathcal{E}_0^{1/3}$ at large $\gamma \mathcal{E}_0$ in the non-equilibrium initial state. However, as shown in Fig. \ref{Fig:TAnisoFE}, we do find that $T_0 \sim \mathcal{E}_0^{1/3}$ at large $\gamma \mathcal{E}_0$ for any typical initial state when the variables of the subsystem shear stress tensors (and $\delta =0$) are fixed. This indicates that the non-equilibrium shear-stress tensor does not affect the metric coupling equations sufficiently to change the conformal scaling of the energy density at large  $\gamma \mathcal{E}_0$ originating from the thermal part of the full energy momentum tensor.

\begin{figure}
 \centering
    \includegraphics[width=0.49\textwidth]{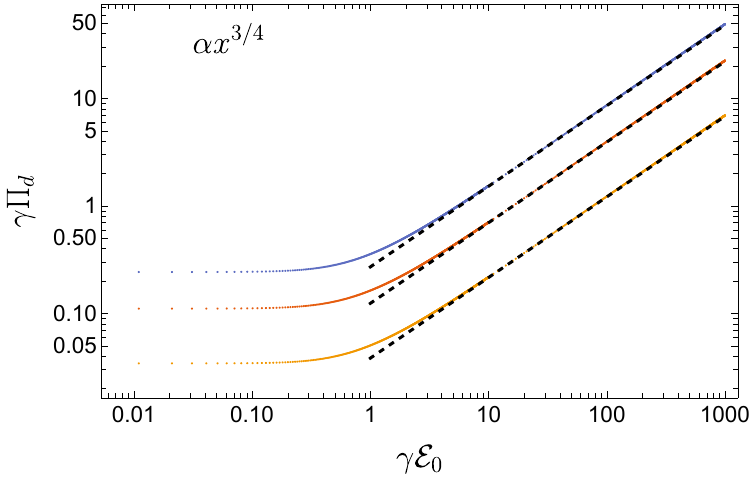}
    \includegraphics[width=0.49\textwidth]{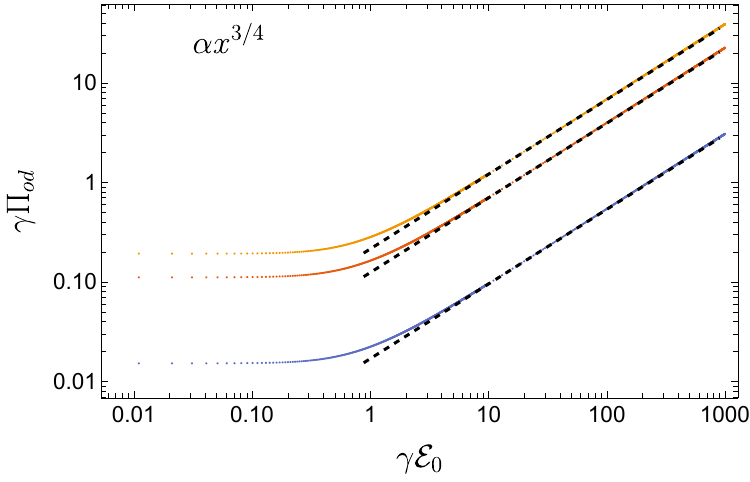}
    \caption{Dependence of the total diagonal anisotropy $\Pi_d = \Pi_{22} = -\Pi_{33}$ (left) and off-diagonal anisotropy $\Pi_{od} = \Pi_{23} = \Pi_{32}$ (right) of the system as a function of the energy density of the full system for different initial anisotropy in \Cref{tab:Tab1}.}
     \label{Fig:TAnisoFE1}
 \end{figure}

Furthermore, in a typical initial state, the diagonal ($\Pi_d$) and off-diagonal ($\Pi_{od}$) parts of the full shear stress tensor also scale with the total energy density as $\sim (\gamma\mathcal{E}_0)^{3/4}$ when $\delta$ and the variables of the subsystem stress tensor are kept fixed as demonstrated in Fig. \ref{Fig:TAnisoFE1}. Although $\delta=0$ in this figure, this result is valid for any choice of $\delta$. Crucially, the full shear stress tensor grows universally with $\gamma\mathcal{E}_0$ at any value of $\delta$ for random choices of the four variables of the subsystem shear stress tensors (with the exception of fine-tuned choices of these variables).

\begin{figure}
\centering
 \includegraphics[width=0.49\textwidth]{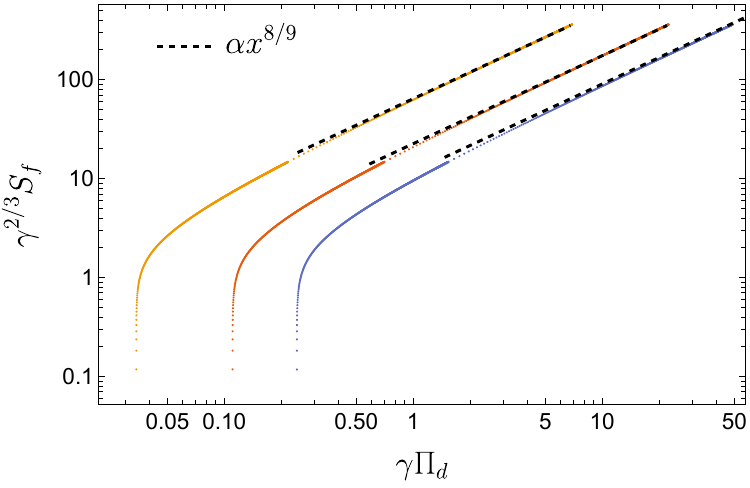}
     \includegraphics[width=0.49\textwidth]{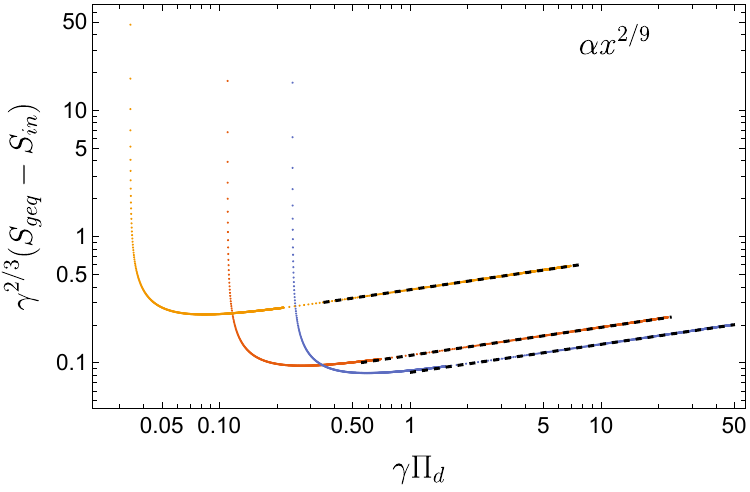}\\
 \includegraphics[width=0.49\textwidth]{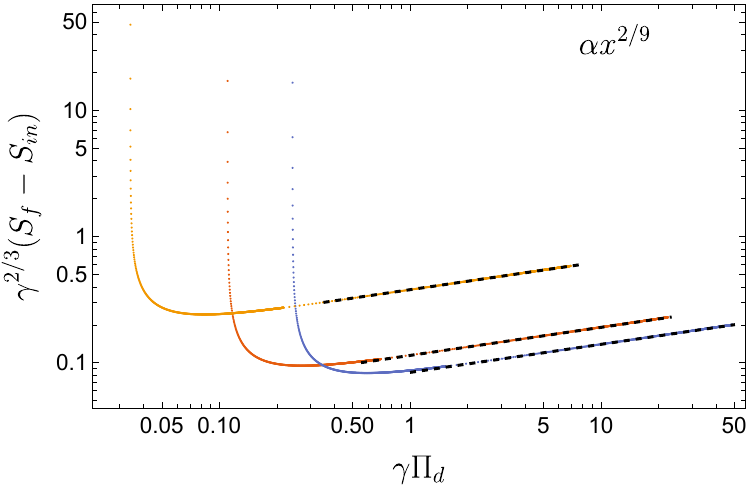}
    \caption{Top left: behavior of the final entropy of the full system as a function $\Pi_d = \Pi_{22} = -\Pi_{33}$. Top right: how far the initial entropy of the system is from the entropy of the global equilibrium state after anisotropy is turned on. Bottom: entropy production as a function of the total anisotropy of the system. The different curves correspond to the anisotropy in \Cref{tab:Tab1}.}
     \label{Fig:EntropyPlot}
 \end{figure}

It can be expected that the entropy of the final pseudo-equilibrium state should have universal characteristics in light of the expectation that the full shear stress tensor governs entropy production and scales with $\mathcal{E}_0$ universally in a typical initial state at large $\gamma \mathcal{E}_0$. \Cref{Fig:EntropyPlot} illustrates the characteristics of the final entropy as a function of the initial anisotropy $\Pi$ (collectively denoting $\Pi_d$ and $\Pi_{od}$) of the full system for different choices of the four variables characterizing the initial subsystem shear stress tensors shown in \Cref{tab:Tab1}. We find these features:
\begin{itemize}
    \item The total entropy density $S_f$ of the final pseudo-equilibrium state scales as $\Pi^{8/9}$ with $\Pi$ denoting $\Pi_d$ or $\Pi_{od}$.
    \item Both the difference between the final entropy and the initial entropy densities $S_f - S_{in}$, and the difference between the global equilibrium entropy and the initial entropy densities $S_{geq} - S_{in}$ at the same value of $\mathcal{E}_0$ scale as $\Pi^{2/9}$ with $\Pi$ denoting $\Pi_d$ or $\Pi_{od}$.
\end{itemize}
These universal scalings hold as $\mathcal{E}_0$ is increased for any initial value of $\delta$ and for typical fixed values of the four variables characterizing the initial subsystem shear stress tensors, although Fig.~\ref{Fig:EntropyPlot} illustrates these for $\delta =0$. The first scaling above is expected only if the final pseudo-equilibrium state is very close to the global equilibrium state at the same value of $\mathcal{E}_0$. Given that $S\sim \mathcal{E}^{2/3}$ at large $\mathcal{E}$ in global thermal equilibrium \cite{Kurkela:2018dku} (recall the discussion at the end of the previous section) and $\Pi \sim \mathcal{E}_0^{3/4}$  (i.e. $\mathcal{E}_0 \sim \Pi^{4/3}$) as discussed above, we expect that $S_f\sim \Pi^{4/3\times 2/3}= \Pi^{8/9}$. This is indeed borne out at large $\gamma\mathcal{E}_0$. 

The equal scalings of  $S_f - S_{in}$ and $S_{geq} - S_{in}$ at large $\mathcal{E}_0$ can also be possible if the final state is very close to the global equilibrium state, i.e. if $$\frac{S_{geq}-S_f}{S_f -S_{in}}\rightarrow 0$$at large $\mathcal{E}_0$. Note that both $S_f - S_{in}$ and $S_{geq} - S_{in}$ decrease with increasing $\Pi$ when $\Pi$ is small. This can be attributed to the negative sign of the square of the shear stress tensor in the BRSSS entropy \eqref{Eq:EntropyRelation}. Only if the total energy is sufficiently large is the entropy production driven by the growth of the thermodynamic components of the entropy.

Finally, we note that $(S_f - S_{in})/(S_{geq} - S_{in})$ goes to unity at large $\mathcal{E}_0$ for typical initial states, i.e. as  $\mathcal{E}_0$ is increased for fixed $\delta$ and fixed random choices of the four variables characterizing the subsystem shear stress tensors of the initial state. Once again, we emphasize that this result holds for an arbitrary choices of $\delta$.

We can conclude that typical initial states of the full isolated system with large energy densities do thermalize after homogeneous thermal relaxation. We want to comment that the emergence of conformality at large $\gamma\mathcal{E}_0$ could play a significant role in the emergence of this universal behavior. It is to be noted that this emergence of conformality in democratic effective metric coupling has been observed also when non-conformal subsystems are coupled via the democratic effective metric coupling \cite{Kurkela:2018dku}, and has also been observed for other types of democratic coupling \cite{Mondkar:2021qsf}. It would be interesting to see if the scalings found at large $\mathcal{E}_0$ could be independent of the microscopic details of the subsystems and can have phenomenological signatures. We leave a more thorough investigation of this issue for the future.

\section{Conclusions}\label{Sec:Conc}

In this work, we have shown that the semi-holographic framework for hybrid system of perturbative and non-perturbative degrees of freedom can provide a consistent paradigm for thermalization in the large $N$ limit. The thermal state, where the subsystems have equal physical temperatures, obeys thermodynamic identities, and the full entropy of the system has the statistical interpretation of the logarithm of the number of microstates. Crucially, this global thermal state is the unique state in the micro-canonical ensemble that has maximum entropy. A typical coarse-grained state in the micro-canonical ensemble relaxes to the global thermal state, and not to a pseudo-equilibrium state, in the large $N$ limit when the average energy density is taken to infinity.

It is interesting to ask if the semi-holographic large $N$ system satisfies the eigenstate thermalization hypothesis \cite{DAlessio:2015qtq}. We cannot examine the expectation values of arbitrary (single-trace) operators in our present investigations, as they are only at the coarse-grained level. Nevertheless, we do find that even at the coarse-grained level, fine-tuned initial conditions can relax to a pseudo-equilibrium state in which the two subsectors have different physical temperatures. This is not inconsistent with the eigenstate thermalization hypothesis, which does rule out atypical behavior when the initial conditions are fine-tuned. 

A better understanding of the quantum statistics of thermalization in large $N$ semi-holography can be obtained from the observation that the large $N$ approximation at the quantum level implies that the time-evolution of the system is effectively a non-linear unitary channel of the form
\begin{equation*}
    U[g_{\mu\nu}]\otimes \tilde{U}[\tilde{g}_{\mu\nu}],
\end{equation*}
where $U[g_{\mu\nu}]$ and $ \tilde{U}[\tilde{g}_{\mu\nu}]$ are the unitary evolution operators constructed with the Hamiltonians of the perturbative and non-perturbative sectors, respectively, in the respective effective background metrics $g_{\mu\nu}$ and $\tilde{g}_{\mu\nu}$. This channel is non-linear because the effective background metrics $g_{\mu\nu}$ and $\tilde{g}_{\mu\nu}$ are determined by $\langle t_{\mu\nu}\rangle$ and $\langle \tilde{t}_{\mu\nu}\rangle$, the expectation values of the subsystem energy-momentum tensors, and thus both $U[g_{\mu\nu}]$ and $ \tilde{U}[\tilde{g}_{\mu\nu}]$ are state-dependent. 

Such a non-linear unitary channel for the perturbative sector with semi-holographic coupling was studied explicitly in \cite{Kibe:2023ixa} (see also \cite{Banerjee:2024ioz}) where a single quantum harmonic oscillator was coupled to a holographic quantum dot described by Jackiw-Teitelboim gravity in the large $N$ limit. In this system, while the entropy of the holographic sector increases, the quantum harmonic oscillator does not lose all of its energy to the black hole. Rather, it goes to a time-dependent state which is determined by the requirement that it decouples from the black hole at late time when the mass of the black hole and the energy of the oscillator become constant. The final time-dependent state also retains a non-isometric copy of the information of the initial state.~\footnote{In other semi-holographic scenarios \cite{Ecker:2018ucc, Mondkar:2021qsf}, where a single field was coupled to holography, the former lost all its energy to the black hole. Interestingly, when a quantum field is coupled semi-holographically to the black hole, the former retains a residual energy and a non-isometric copy of the initial information. It was shown in \cite{Kibe:2023ixa} that another copy of the information of the initial state of the quantum oscillator is transferred to the non-linearities of the ringdown process as the black hole mass settles down to a constant value.} 

For a more precise understanding of the quantum statistics of thermalization in large $N$ semi-holography, we need to understand the dynamics of $\mathcal{O}(N^2)$ quantum oscillators coupled to the black hole via the democratic coupling. {In the presence of the effective metric coupling, we expect that only this coupling survives at late time giving an evolution described by the $U[g_{\mu\nu}]\otimes \tilde{U}[\tilde{g}_{\mu\nu}]$ non-linear unitary channel.} It would be interesting to see how the time evolution captures features predicted by the eigenstate thermalization hypothesis via explicit calculations of time-dependent expectation values of operators.\footnote{For a recent demonstration of how holographic systems realize statistical features of the eigenstate thermalization hypothesis, see \cite{Gannouji:2025nmh}.}  We leave this for the future.

Furthermore, we want to comment that although we have focused on homogeneous relaxation, we expect that the conclusions should hold for arbitrary processes. One can argue that the full isolated system thermalizes locally before undergoing hydrodynamic evolution to a homogeneous global equilibrium. It would be interesting to study this explicitly in the future as well. 

Finally, for phenomenological applications, it would be useful to see if the scaling behavior of the entropy production reported in this work for conformal subsystems holds independently of the choice of the equations of state of the subsystems. It is necessary to investigate this also in the full semi-holographic setup in which a perturbative quantum gas of particles is coupled to a holographic geometry representing the non-perturbative sector via democratic effective metric coupling.

\acknowledgments

The authors would like to thank Anton Rebhan for useful discussions. TM has been supported by an appointment to the JRG Program at the APCTP
through the Science and Technology Promotion Fund and Lottery Fund of the Korean Government, by the Korean
Local Governments – Gyeongsangbuk-do Province and Pohang City -  and by NRF funded by the Korean government (MSIT) (grant number 2021R1A2C1010834). TM has been supported by the DFG through the Emmy Noether Programme (project
number 496831614), through CRC 1225 ISOQUANT (project number 27381115). AM is supported by Fondecyt grant 1240955. AS is supported by funding from Horizon Europe research and innovation programme under the Marie Sk\l odowska-Curie grant agreement No.~101103006. 

\appendix

\section{Explicit expressions}\label{Sec:Lengthy}

The full determinant discussed around \cref{det=0} is given by
\begin{align}\label{det=0full}
    \text{det} (Q) &=  \gamma ^{5/3} \Big(a(\lambda )
   \temp(\lambda )-\At(\lambda ) \tempt(\lambda )\Big) \frac{4  b(\lambda ) \Bt(\lambda ) \Big(P(\lambda )+\e (\lambda )\Big) \Big(\pt(\lambda )+\et(\lambda )\Big) }{\At(\lambda ) \cs(\lambda )^2 \cst(\lambda )^2 \temp(\lambda )^2
   \tempt(\lambda )^2}  \nonumber \\ 
   & \Bigg( -4 a(\lambda )^3 \At(\lambda )^3 b(\lambda )^2
   \Bt(\lambda )^2 
   + \gamma ^2 \Big(4 a(\lambda )^2
   \At(\lambda )^2 b(\lambda )^2 \Bt(\lambda )^2 P(\lambda ) \pt(\lambda ) \nonumber \\
   &+4 a(\lambda
   )^4 \At(\lambda )^4 \cs(\lambda )^2 \cst(\lambda )^2 (P(\lambda )+\e (\lambda
   )) (\pt(\lambda )+\et(\lambda ))+b(\lambda )^4 \Bt(\lambda )^4 \e
   (\lambda ) \et(\lambda ) \nonumber \\
   &+r \left(4 a(\lambda )^2 b(\lambda )^2
   P(\lambda ) \left(-3 \At(\lambda )^2 \Bt(\lambda )^2 \pt(\lambda )+2
   \At(\lambda )^4 \cst(\lambda )^2 (\pt(\lambda )+\et(\lambda
   ))-\Bt(\lambda )^4 \et(\lambda )\right) \right.\nonumber \\& \left. +8 a(\lambda )^4 \At(\lambda )^2
   \cs(\lambda )^2 (P(\lambda )+\e (\lambda )) \left(\Bt(\lambda )^2 \pt(\lambda
   )-2 \At(\lambda )^2 \cst(\lambda )^2 (\pt(\lambda )+\et(\lambda
   ))\right) \right. \nonumber \\& \left. -2 b(\lambda )^4 \Bt(\lambda )^2 \e (\lambda ) \left(2 \At(\lambda )^2
   \pt(\lambda )+\Bt(\lambda )^2 \et(\lambda )\right)\right) \nonumber \\& + r^2 \left(-4 a(\lambda )^2 b(\lambda )^2
   P(\lambda )+4 a(\lambda )^4 \cs(\lambda )^2 (P(\lambda )+\e (\lambda ))-b(\lambda )^4
   \e (\lambda )\right) \nonumber
   \\   
   &\left(-4 \At(\lambda )^2 \Bt(\lambda )^2 \pt(\lambda )+4
   \At(\lambda )^4 \cst(\lambda )^2 (\pt(\lambda )+\et(\lambda
   ))-\Bt(\lambda )^4 \et(\lambda )\right)\Big) \nonumber \\ 
   & + \gamma ^4 \Big(-a(\lambda ) \At(\lambda ) b(\lambda
   )^2 \Bt(\lambda )^2 \left(\cs(\lambda )^2 P(\lambda ) \e (\lambda
   )+\cs(\lambda )^2 \e (\lambda )^2+P(\lambda )^2\right)  \nonumber \\ &  \left(\cst(\lambda )^2
   \pt(\lambda ) \et(\lambda )+\cst(\lambda )^2 \et(\lambda
   )^2+\pt(\lambda )^2\right)\nonumber \\ & +6 r a(\lambda )
   \At(\lambda ) b(\lambda )^2 \Bt(\lambda )^2 \left(\cs(\lambda )^2 P(\lambda )
   \e (\lambda )+\cs(\lambda )^2 \e (\lambda )^2+P(\lambda )^2\right) \nonumber    \end{align}
   \begin{align}%\\ 
   & 
   \left(\cst(\lambda )^2 \pt(\lambda ) \et(\lambda )+\cst(\lambda )^2
   \et(\lambda )^2+\pt(\lambda )^2\right) \nonumber \\ & -9 r^2 a(\lambda ) \At(\lambda ) b(\lambda )^2 \Bt(\lambda )^2
   \left(\cs(\lambda )^2 P(\lambda ) \e (\lambda )+\cs(\lambda )^2 \e (\lambda
   )^2+P(\lambda )^2\right)  \nonumber \\ & \left(\cst(\lambda )^2 \pt(\lambda ) \et(\lambda
   )+\cst(\lambda )^2 \et(\lambda )^2+\pt(\lambda )^2\right)\Big) \Bigg)
\end{align}

Below, we give the explicit evolution equations of the hybrid {BRSSS} system.
The Ward identity for the conservation of energy and momentum of the first sector is given by the following equation
\begin{align}\label{ward-1-explicit}
    0 &=  6
  \e'(t)  \A_{11}(t) \A_{33}(t)  \Big(\A_{22}(t)^2
\A_{33}(t)^2-\A_{23}(t)^2\Big) \nonumber \\  &+ \e(t) \A_{33}(t)  \Big(\A_{22}(t) \A_{33}(t)
   \left(\A_{22}(t) \left(3 \A_{11}(t) \A_{33}'(t)-8 \A_{33}(t)
   \A_{11}'(t)\right)+3 \A_{11}(t) \A_{33}(t) \A_{22}'(t)\right)  \nonumber \\  &  +8
   \A_{23}(t)^2 \A_{11}'(t)-3 \A_{11}(t) \A_{23}(t)
   \A_{23}'(t)\Big) \nonumber \\ &+ \py_d(t)  \A_{11}(t) \Big(-\A_{22}(t)
\A_{33}(t)^3 \A_{22}'(t)+\A_{22}(t)^2 \A_{33}(t)^2
\A_{33}'(t)+\A_{23}(t) \left(\A_{33}(t) \A_{23}'(t)-2 \A_{23}(t)
\A_{33}'(t)\right)\Big) \nonumber \\ & -2 \py_{od}(t) \A_{11}(t) \A_{22}(t) 
   \Big(\A_{23}(t) \left(\A_{33}(t) \A_{22}'(t)+\A_{22}(t)
\A_{33}'(t)\right)-\A_{22}(t) \A_{33}(t) \A_{23}'(t)\Big)
\end{align}
for homogeneous evolution. The BRSSS equations \eqref{Eq:chap6mis} for the first subsector is given explicitly by
\begin{align}\label{mis-11-explicit}
    0 &=  \py_d'(t) \Big(\tp
   \A_{23}(t)^2 \A_{33}(t)^2-\tp \A_{22}(t)^2
   \A_{33}(t)^4\Big) +\py_{od}'(t) \Big(2 \tp
   \A_{23}(t)^3-2 \tp \A_{22}(t)^2 \A_{23}(t)
   \A_{33}(t)^2\Big) \nonumber \\ & + \py_d(t)  \Big( \A_{11}(t) \A_{33}(t)^2  \left(\A_{23}(t)^2-\A_{22}(t)^2
   \A_{33}(t)^2\right) \Big) + 2 \py_{od}(t) \Big( \A_{11}(t) \A_{23}(t) 
   \left(\A_{23}(t)^2-\A_{22}(t)^2 \A_{33}(t)^2\right) \Big) \nonumber \\&+ \eta  \Big(2
   \A_{22}(t) \A_{33}(t)^4 \A_{22}'(t)-2 \A_{22}(t)^2 \A_{33}(t)^3
   \A_{33}'(t)-2 \A_{23}(t) \A_{33}(t)^2 \A_{23}'(t)+4 \A_{23}(t)^2
   \A_{33}(t) \A_{33}'(t)\Big),
   \end{align}
   \begin{align}
\label{mis-12-explicit}
    0 &=   \py_d'(t) \Big(\tp
   \A_{22}(t)^2 \A_{23}(t) \A_{33}(t)^3-\tp \A_{23}(t)^3
   \A_{33}(t)\Big) +\py_{od}'(t) \Big(2 \tp \A_{22}(t)^4
   \A_{33}(t)^3-2 \tp \A_{22}(t)^2 \A_{23}(t)^2
   \A_{33}(t)\Big) \nonumber \\ & +  \py_{od}(t) \Big(2 \A_{11}(t) \A_{22}(t)^2 \A_{33}(t)
   \left(\A_{22}(t)^2 \A_{33}(t)^2-\A_{23}(t)^2\right)-2 \tp
   \A_{22}(t) \A_{23}(t)^2 \A_{33}(t) \A_{22}'(t) \nonumber \\ & +2 \tp
   \A_{22}(t)^3 \A_{33}(t)^3 \A_{22}'(t)+2 \tp \A_{22}(t)^2
   \A_{23}(t)^2 \A_{33}'(t)-2 \tp \A_{22}(t)^4 \A_{33}(t)^2
   \A_{33}'(t)\Big) \nonumber \\ & +\py_d(t) \Big(\A_{11}(t) \A_{23}(t)
   \A_{33}(t) \left(\A_{22}(t)^2
   \A_{33}(t)^2-\A_{23}(t)^2\right)+\tp \A_{22}(t)^2
   \A_{33}(t)^3 \A_{23}'(t) \nonumber \\ &-2 \tp \A_{22}(t)^2 \A_{23}(t)
   \A_{33}(t)^2 \A_{33}'(t)-\tp \A_{23}(t)^2 \A_{33}(t)
   \A_{23}'(t)+2 \tp \A_{23}(t)^3 \A_{33}'(t)\Big)\nonumber \\ & +\eta 
   \Big(-2 \A_{22}(t) \A_{23}(t) \A_{33}(t)^3 \A_{22}'(t)+2
   \A_{22}(t)^2 \A_{33}(t)^3 \A_{23}'(t)-2 \A_{22}(t)^2 \A_{23}(t)
   \A_{33}(t)^2 \A_{33}'(t)\Big)
\end{align}
For the other subsector, we need to replace all variables and parameters by the corresponding ones in the other subsystem (tilde $\leftrightarrow$ untilded)  to obtain the corresponding set of equations, i.e. Ward identity and the BRSSS equations.
The coupling equations are:
\begin{align}\label{coupling-explicit}
    0 &=  -1 + \A_{11}(t)^2+\At_{11}(t) \sqrt{\At_{22}(t)^2
   \At_{33}(t)^2-\At_{23}(t)^2} \Big(\gamma  \frac{
   \et(t)}{\At_{11}(t)^2}  + \gamma  r \Big(-\frac{\et(t)}{\At_{11}(t)^2} \nonumber \\& +\frac{\At_{33}(t)^2 ((t)-\pyt_d(t))-2 \At_{23}(t) \pyt_{od}(t)}{2 \At_{22}(t)^2
   \At_{33}(t)^2-2 \At_{23}(t)^2} \nonumber \\&   +\frac{\At_{22}(t)^2 \left(\At_{33}(t)^2
   (\pyt_{d}(t)+\et(t))-2 \At_{23}(t) \pyt_{od}(t)\right)-2 \At_{23}(t)^2 \pyt_{d}(t)}{2 \At_{22}(t)^2
   \At_{33}(t)^4-2 \At_{23}(t)^2 \At_{33}(t)^2}\Big)\Big) \\ 
   0 &=   1-\A_{22}(t)^2+\At_{11}(t) \sqrt{\At_{22}(t)^2
   \At_{33}(t)^2-\At_{23}(t)^2} \Big(\gamma \frac{  \left(\At_{33}(t)^2
   (\et(t)-\pyt_{d}(t))-2 \At_{23}(t) \pyt_{od}(t)\right)}{2 \At_{22}(t)^2 \At_{33}(t)^2-2 \At_{23}(t)^2}  \nonumber \\ & -\gamma  r
   \Big(-\frac{\et(t)}{\At_{11}(t)^2}+\frac{\At_{33}(t)^2
   (\et(t)-\pyt_{d}(t))-2 \At_{23}(t) \pyt_{od}(t)}{2
   \At_{22}(t)^2 \At_{33}(t)^2-2 \At_{23}(t)^2}\nonumber \\ & +\frac{\At_{22}(t)^2
   \left(\At_{33}(t)^2 (\pyt_{d}(t)+\et(t))-2 \At_{23}(t)
   \pyt_{od}(t)\right)-2 \At_{23}(t)^2 \pyt_{d}(t)}{2
   \At_{22}(t)^2 \At_{33}(t)^4-2 \At_{23}(t)^2
   \At_{33}(t)^2}\Big)\Big) \\
   0 &=  1 -\A_{33}(t)^2 \nonumber \\ & +\At_{11}(t) \sqrt{\At_{22}(t)^2
   \At_{33}(t)^2-\At_{23}(t)^2} \Big(\gamma \frac{  \left(\At_{22}(t)^2
   \left(\At_{33}(t)^2 (\pyt_{d}(t)+\et(t))-2 \At_{23}(t)
   \pyt_{od}(t)\right)-2 \At_{23}(t)^2 \pyt_{d}(t)\right)}{2
   \At_{22}(t)^2 \At_{33}(t)^4-2 \At_{23}(t)^2 \At_{33}(t)^2} \nonumber \\ &- \gamma  r
   \Big(-\frac{\et(t)}{\At_{11}(t)^2}+\frac{\At_{33}(t)^2
   (\et(t)-\pyt_{d}(t))-2 \At_{23}(t) \pyt_{od}(t)}{2
   \At_{22}(t)^2 \At_{33}(t)^2-2 \At_{23}(t)^2} \nonumber \\
   & +\frac{\At_{22}(t)^2
   \left(\At_{33}(t)^2 (\pyt_{d}(t)+\et(t))-2 \At_{23}(t)
   \pyt_{od}(t)\right)-2 \At_{23}(t)^2 \pyt_{d}(t)}{2
   \At_{22}(t)^2 \At_{33}(t)^4-2 \At_{23}(t)^2
   \At_{33}(t)^2}\Big)\Big) \\
   0 &=  -\A_{23}(t) + \gamma \frac{  \At_{11}(t) \sqrt{\At_{22}(t)^2 \At_{33}(t)^2-\At_{23}(t)^2}
   \left(2 \At_{22}(t)^2 \pyt_{od}(t)+\At_{23}(t) (\pyt_d(t)-\et(t))\right)}{2 \At_{22}(t)^2 \At_{33}(t)^2-2
   \At_{23}(t)^2}
   \end{align}

\bibliographystyle{jhep}
\bibliography{draft}

\end{document}